\documentclass[12pt]{amsart}

\usepackage{amsfonts}
\usepackage[all]{xy} 
\usepackage{graphicx} 
\usepackage{color} 

\newcommand{\vc}[1]{\boldsymbol{#1}} 
\newcommand{\disk}{\ensuremath{\Delta} } 
\newcommand{\cdisk}{\ensuremath{\overline{\Delta}}} 
\newcommand{\rqt}{\widetilde{T}^{B}_{\#}} 
\newcommand{\Chat}{\hat{\Bbb{C}}} 
\newcommand{\riem}[1][]{\ensuremath{\Sigma_{#1}}}  
\newcommand{\pmcgi}[1][]{\mathrm{PModI}({#1})} 
\renewcommand{\Bbb}[1]{\ensuremath{\mathbb{#1}}}
\newcommand{\st}{\, | \,} 
\newcommand{\ann}{\Bbb{A}}  

\theoremstyle{plain}
        \newtheorem{theorem}{Theorem}[section]
        \newtheorem{lemma}[theorem]{Lemma}
        \newtheorem{proposition}[theorem]{Proposition}
        \newtheorem{corollary}[theorem]{Corollary}

\theoremstyle{definition}
        \newtheorem{definition}[theorem]{Definition}

\theoremstyle{remark}
    \newtheorem{remark}[theorem]{Remark}

\numberwithin{equation}{section} 
\numberwithin{figure}{section} 

\begin{document}


\title{Quasisymmetric sewing in rigged Teichm\"uller space}

\author{David Radnell}

\date{October 13, 2005}

\address{ Department of Mathematics \\
University of Michigan \\
Ann Arbor, MI 48109-1043, USA}
\email[D. ~Radnell]{radnell@umich.edu}

\author{Eric Schippers}
\address{Department of Mathematics \\
University of Manitoba\\
Winnipeg, MB, Canada, R3T 2N2}
\email[E. ~Schippers]{Eric\_Schippers@UManitoba.CA}

\begin{abstract}
One of the basic geometric objects in conformal field theory (CFT)
is the  the moduli space of Riemann surfaces whose $n$ boundaries
are ``rigged'' with analytic parametrizations. The fundamental
operation is the sewing of such surfaces using the
parametrizations to identify points.  An alternative model is the
moduli space of $n$-punctured Riemann surfaces together with local
biholomorphic coordinates at the punctures.  We refer to both of
these moduli spaces as the ``rigged Riemann moduli space".

By generalizing to quasisymmetric
boundary parametrizations, and defining rigged Teichm\"uller
spaces in both the border and puncture pictures, we prove the
following results:
(1) The Teichm\"uller space of a genus-$g$ surface bordered by
$n$ closed curves covers the  rigged
Riemann and rigged Teichm\"uller moduli spaces of surfaces of the
same type, and induces complex manifold structures on them.  (2)
With this complex structure the sewing operation is holomorphic.
(3) The border and puncture pictures of the rigged moduli and
rigged Teichm\"uller spaces are biholomorphically equivalent.

These results are necessary in rigorously
defining CFT (in the sense of G. Segal), as well as for the construction of
CFT from vertex operator algebras.

\end{abstract}

\keywords{Teichm\"uller spaces, conformal field theory, sewing,
rigged Riemann surfaces} \subjclass[2000]{Primary 30F60; Secondary
81T40}

\maketitle

\tableofcontents


\begin{section}{Introduction}
\label{introduction}


\begin{subsection}{Statement and discussion of results}

This paper is devoted to the rigorous construction of some of the
fundamental objects and operations in the geometric approach to
two-dimensional conformal field theory.  The central object is the
\textit{rigged Riemann surface}.  We use this term to refer to both:
(1) a bordered Riemann surface with parametrizations of its boundary
components by the unit circle $S^1$, and (2) a punctured surface with
specified local biholomorphic coordinates at the punctures.  The space
of conformal equivalence classes of rigged Riemann surfaces is called
the \textit{rigged moduli space}.

Our main results are as follows.  Let $T^B(g,n)$ denote the infinite-dimensional
Teichm\"uller space of Riemann surfaces of genus $g$ bounded by $n$ closed curves.
We construct natural ``rigged Teichm\"uller spaces" for both the border and
puncture model, denoted $\widetilde{T}_\#^B(g,n)$ and $\widetilde{T}^P(g,n)$
respectively, which cover the rigged Riemann moduli space.
\begin{enumerate}
\item The rigged Riemann moduli space and the rigged Teichm\"uller
space are obtained as a quotients of $T^B(g,n)$ by discrete
subgroups of the mapping class group.
In particular they inherit a natural complex structure from
$T^B(g,n)$.  An essential ingredient of these constructions is the use
of the class of quasisymmetric boundary parametrizations rather than
the standard analytic ones.
\item Once this complex structure is constructed, we define the
sewing operation for
quasisymmetric riggings and show that it is holomorphic (Theorem
\ref{th:sewingisholo}).
\item The puncture and border models of both the rigged Riemann moduli space
and the rigged Teichm\"uller space are biholomorphically equivalent.
\end{enumerate}


As far as we are aware, this is the first published proof that the
sewing operation is holomorphic for arbitrary genus, even in the
analytic case. In the genus-zero case, the rigorous construction of the
(analytically) rigged moduli space was first given by Huang in
\cite{Huang}, and the holomorphicity of the sewing was proved using
a theorem of Fischer and Grauert.
Part of the purpose of this paper is to prove the
higher-genus versions of these results.

The higher-genus case with analytic parametrizations was completed in
the first author's thesis \cite{Radnell_thesis}, but is yet to be
published. The methods and constructions in \cite{Radnell_thesis} are
of a conceptually different nature to those of the present paper. In
fact the results are complementary. The relation between the two
approaches is outlined in Section \ref{se:compat_thesis} and will be
the subject of a future publication.

The generalization from analytic to quasisymmetric riggings is both
natural and necessary from the point of view of Teichm\"uller theory,
since quasisymmetric mappings are the boundary values of
quasiconformal maps.  In fact, this generalization is essential to
showing that the Teichm\"uller space $T^B(g,n)$ covers the rigged
Riemann moduli space, in part by making technical tools available, and
also by providing a clearer framework.  It also made it possible to
show that the puncture and border models are biholomorphically
equivalent on both the Riemann and Teichm\"uller space levels.  This
was not previously possible and allows us to fill out a conceptually
satisfying commutative diagram (see Diagram \eqref{house}).

Moreover, our generalization to quasisymmetric boundary
parametrizations has applications in the construction of CFT from
vertex operator algebras.  This is explained at the end of Section
\ref{background}.  Unexpectedly, it also gives an apparently new
description of the Teichm\"uller space of a bordered Riemann
surface as a cover of the rigged Teichm\"uller space.  Some
results on the local structure of rigged Teichm\"uller space are
given in Section \ref{se:localstructure}, and a possible
application to describing $T^B(g,n)$ is given in the conclusion.

The motivation for the results obtained in this paper is as follows (a
more detailed historical background is given in Section
\ref{background}.)  In Segal's foundational work
\cite{SegalPublished}, rigged Riemann surfaces were introduced and a
mathematical definition of conformal field theory was proposed (as
outlined in Section \ref{background}). Rigorous construction of CFT
in Segal's sense is a major mathematical undertaking. Significant
progress has been made by Huang and others by developing and using the
theory of vertex operator algebras
\cite{Huang_91,Huang_97,Huang_Diff_Eq, Huang_Diff_Int} and \cite{Zhu}.
Indeed, construction of genus-zero theories is now complete and
genus-one is essentially complete.  The results of this paper are a
necessary part of Huang's program to construct higher-genus theories.
A separate issue in rigorizing the definition of CFT involves the
categorical structure of rigged surfaces (together with the operations
of sewing and disjoint union). Significant recent work by Hu, Kriz and Fiore
in \cite{Hu_Kriz04, Hu_Kriz05} and \cite{Fiore} 
has successfully dealt with this issue.

The idea of applying constructions in geometric function theory to
conformal field theory and string theory is not new
(e.g. \cite{Bowick_Rajeev}, \cite{Lempert}, \cite{NagandVerjovsky},
\cite{Pekonen}, \cite{Takhtajan_TeoI} and \cite{Takhtajan_TeoII}).  It
has been suggested (see the comprehensive review of Pekonen
\cite{Pekonen} and references therein) that the universal
Teichm\"uller space would provide the natural arena for conformal
field theory, and the results of this paper indicate that this is in
fact correct.

We also hope that this paper will encourage more interaction
between conformal field theory and geometric function theory. To
further this goal we have outlined in Section
\ref{se:compat_thesis} the relation of our model to the standard
conformal field theory model involving punctured surfaces and
analytic local coordinates. This problem in itself is not entirely
trivial and the full details will appear in a forthcoming article.

\end{subsection}


\begin{subsection}{Background} \label{background}
In this section we motivate the paper from the
point of view of conformal field theory (CFT).

A complete understanding of CFT requires
geometry and analysis as well as algebra. In fact, even making the definition of CFT
rigorous is highly non-trivial.
This paper can be seen as part of the program of
constructing the geometric
objects and operations in CFT. The
results will have applications to the program of rigorously
constructing CFT from vertex operator algebras. We briefly discuss
the basic concepts of CFT in order to provide context.

Conformal field theory originally arose in physics from various
two-dimensional statistical mechanics models. In the seminal
paper of Belavin, Polyakov and Zamolodchikov \cite{BPZ} much of
the structure of CFT was encoded in the notion of a chiral
algebra, at the physical level of rigor. These algebras are
essentially equivalent to vertex operator algebras which were
developed independently in mathematics by Borcherds \cite{Bor} and
Frenkel, Lepowsky and Meurman \cite{FLM}.

At around the same time in string theory the study of the geometry
of CFT was introduced by Friedan and Shenker \cite{FS}. In this
context the two dimensional objects of study are the world sheets
of strings which  are Riemann surfaces with boundary.

In the path integral approach to quantum field theory, one must
``sum'' over all possible paths. In the case of string
interactions the possible paths are the possible Riemann surfaces
joining the prescribed boundaries (strings). The conformal
invariance inherent in the physics requires that the path
integrals be taken over the moduli space. However, it is doubted that these
path integrals can be made rigorous in a direct way.

Around 1987, Segal
\cite{Segal87, Segal88, SegalPublished} and Kontsevich
independently extracted the mathematical properties such a theory
should have and gave a purely mathematical definition of
CFT. Substantial work was done recently by Hu, Kriz and Fiore in
\cite{Hu_Kriz04, Hu_Kriz05} and \cite{Fiore} to rigorize the categorical
structures in this definition. Details of the analytic aspects have
been worked out by Radnell \cite{Radnell_thesis} and will appear in 
a forthcoming article.

In CFT each boundary circle must be associated with a
Hilbert space and each interaction (Riemann surface with parametrized boundary components)
must be associated with an operator.  Such Riemann surfaces can be sewn using
the boundary parametrizations and certain geometric properties of this
operation translate into relations between the corresponding operators.

Collecting these ideas we present an outline of the definition of
a \emph{conformal field theory} in the sense of
Segal (see \cite{SegalPublished} or the review article
\cite{Huang_CFT}). Consider the category, $\mathcal{C}$, whose
objects are ordered sets of copies of the unit circle $S^1$ and
whose morphisms are conformal equivalence classes of Riemann
surfaces with oriented, ordered, and analytically parametrized
boundaries such that the negatively (positively) oriented
boundaries are parametrized by the copies of $S^1$ in the domain
(co-domain). Composition of morphisms is defined by the sewing of
oppositely oriented boundary components in the unique way
specified by the parametrizations. A conformal field theory is a
projective functor from this category to the category of complete
locally convex vector spaces over $\mathbb{C}$, satisfying certain
natural axioms.

Although this definition has existed since 1987, no general
construction for arbitrary genus has been given. This attests to the
richness of the mathematical structure of CFT and the difficulties
faced in its construction.  The genus-zero theory has been completely
worked out by Huang in, \cite{Huang_91}, \cite{Huang_97},
\cite{Huang}, \cite{Huang_99}, and \cite{Huang_Diff_Int}.  The genus-one
theory is also essentially complete due to the work of Zhu \cite{Zhu}
and Huang \cite{Huang_Diff_Eq}.  Free fermion theories were outlined
by Segal \cite{SegalPublished} and have recently been elaborated on by
Kriz \cite{Kriz03}.  Many people have worked on the algebro-geometric
and topological aspects of the higher-genus theory. Some key works are
in this direction are \cite{Kirillov},
\cite{BD}, \cite{BDM},
\cite{FB-Z}, \cite{Gaitsgory}, \cite{HL_Dmod}, \cite{Nagatomo_Tsuchiya},
\cite{TUY}, and \cite{Turaev}.

To construct CFT completely however, many holomorphicity issues must
be addressed.  A richer mathemtical structure is contained in the
chiral and anti-chiral parts of CFT, and in the construction of CFT
from vertex operator algebras it is actually these parts which are
constructed first.  Axiomatically such structures are \textit{weakly
conformal field theories} as defined by Segal \cite{SegalPublished}.
In the chiral case, the operators in the CFT are required to depend
holomorphically on the associated Riemann surface with parametrized
boundary components. For such a statement to make sense the moduli
space of Riemann surfaces with parametrized boundaries must be a
complex manifold and the sewing operation is required to be
holomorphic.

A particular problem in higher-genus CFT is completing the modules
of vertex operator algebras to obtain the required Hilbert spaces
and constructing trace-class maps, associated to Riemann surfaces
with parametrized boundaries, between tensor powers of the Hilbert
spaces. In this completion process and the construction of the
maps, being able to perform the sewing operation with
parametrizations that are more general than analytic is
necessary. Our generalization to quasisymmetric boundary
parametrizations in this paper was partly motivated by this
application. It is crucial that the sewing operation be
holomorphic, and we prove this fact in Section \ref{sewing}.

\end{subsection}


\begin{subsection}{Outline}
In order that this paper be accessible to both those working in
conformal field theory and geometric function theory,  brief
sketches of material considered standard are provided to make the
paper self-contained.  Also,  although all definitions are given
in the main body of the paper, a partial list of notation appears
in Section \ref{notation} for the convenience of the reader.

To get a conceptual picture of the results as quickly as possible,
it may be best to look at Diagram \eqref{house} once the appropriate
definitions have been absorbed. The holomorphicity of the sewing
is indicated in Diagram \eqref{di:sew}, and this can be understood
once the sewing operation from Section \ref{sewing} is read along
with Theorems \ref{TBcoversMB} and \ref{th:complex_moduli}.

An outline of the contents of the paper follows.

Section \ref{preliminaries} contains some basic concepts,
definitions and standard results. We present the definitions and
background on mapping class groups, Teichm\"uller theory, the
extended $\lambda$-lemma and quasisymmetric maps (with some
extensions of the standard terminology).
 In Section \ref{sewing}, we generalize the `conformal welding' construction of geometric function
theory in order to sew general Riemann surfaces with quasisymmetric boundary parametrization.

The main technical results appear in Section \ref{qs_extensions}.
We derive various lemmas regarding the construction of
quasiconformal mappings with specified properties.  These lemmas
are first applied to give the relation between the mapping class
groups of bordered Riemann surfaces and the mapping class groups
of the punctured Riemann surfaces obtained by sewing on `caps'
(that is, copies of the disk).  Although this relation has been
given in the case of the homeomorphic and diffeomorphic setting,
it does not seem to exist in the quasiconformal setting. The
technical lemmas of this section are also crucial to proving the
relation between the Teichm\"uller and Riemann moduli spaces of
the border and puncture model.

Sections \ref{ModuliandTeichmuller} and \ref{se:holo_sewing} contain
 the main results.  In Section \ref{ModuliandTeichmuller}, we present
 the definitions of the of the rigged Teichm\"uller and Riemann moduli
 spaces of bordered and punctured surfaces.  The relation between all
 the spaces and the construction of their complex structures is given.
 These results are summarized in Diagram \eqref{house} and Theorem
 \ref{th:resultssummary}. Once the complex structures are constructed,
 it is proved that the sewing operation is holomorphic in Section
 \ref{se:holo_sewing}.  This is the content of Theorem
 \ref{th:sewingisholo}.

In Section \ref{se:localstructure} we take an important step towards
understanding the local (fiber) structure of the rigged Teichm\"uller
space. In particular we prove in Corollary \ref{r_holo} that a
holomorphic family of riggings gives a holomorphic family in the
rigged Teichm\"uller space.

Although the results of this paper clearly indicate that the
quasisymmetric approach to the riggings is both natural and necessary,
it is of interest to establish the exact relation between this new
approach and the standard analytic rigged moduli space.  A sketch of
the relation is given in Section \ref{se:compat_thesis}, with details
to appear in a later publication.  In particular the compatibility of
the complex structure on the space of germs of holomorphic functions
with that on rigged Teichm\"uller space is demonstrated.

Finally, Section \ref{se:conclusion} contains some concluding remarks,
and a notation key is provided in Section \ref{notation}.

\end{subsection}

\end{section}


\begin{section}{Preliminaries}
\label{preliminaries}

Let $n^-$ and $n^+$ be non-negative integers and let $n = n^- +
n^+$.  An \textit{oriented point} on a Riemann surface is a point
together with an element of $\{+, -\}$. Let $\riem$ be a compact
Riemann surface of genus $g$, and choose a set of ordered,
oriented and distinct points $\mathbf{p} =(p_1,\ldots,p_{n})$
where $n^-$ points are negatively oriented and $n^+$ points are
positively oriented. Let $\riem^P = \riem \setminus \mathbf{p}$ be
the corresponding punctured surface. We say $\riem^P$ is of type
$(g,n^-,n^+)$.  Where it causes no confusion we will sometimes
think of $\riem^P$ as a surface with marked points rather than
punctures.

Let $\riem^B$ be a Riemann surface bounded by $n$ closed curves
which are homeomorphic to $S^1$ and such that sewing in $n$ disks
would result in a compact Riemann surface of genus $g$.  We assign
an order to the set of boundary components and assign an element
of $\{+, -\}$ to each boundary component such that $n^-$
components are negative and $n^+$ are positive.  We will also say
$\riem^B$ is of type $(g,n^-,n^+)$ in this case. We denote the
boundary of $\riem^B$ by $\partial \riem^B$ and the  $i$th
boundary component by $\partial_i \riem^B$. When we write
$\partial \riem^B$ for $ \partial_1 \riem^B \cup \cdots \cup
\partial_{n} \riem^B$ an ordering of the components is implicit.

The orientation is often not important and we will simply refer to the
surface as having $n$ punctures or boundary components.

\begin{remark}
\label{re:inout}
Each boundary curve has an \textit{orientation} which is the
(topological) orientation induced from the Riemann surface
structure. To avoid a conflict in terminology we will refer to
boundary components with an assignment of `$-$' (respectively, `$+$')
as \textit{incoming} (respectively, \textit{outgoing}).  For the
relation to conformal field theory and boundary parametrizations see
Remark \ref{re:CFTorientation}.
\end{remark}


\begin{subsection}{Mapping class groups}
\label{se:mcg}

When dealing with surfaces with punctures or boundary there are
several different mapping class (or modular) groups that can be
considered, depending on how the boundaries and punctures are to be
preserved under homeomorphisms and homotopies.  In this section some
standard definitions and results are presented. As is natural in
Teichm\"uller theory we work solely with quasiconformal
homeomorphisms.  The boundary curves and punctures will always be
ordered and all maps are required to preserve the given ordering.
Some general sources for this material are \cite{Birman},
\cite{Ivanov}, \cite{Nag} and \cite{ParisRolfsen}.

 \begin{definition}
  For a Riemann surface $\riem^B$ bounded by an ordered set of $n$
  curves, let $\mathrm{PQC}^B(\riem^B)$ be the space of quasiconformal
  self-mappings of $\riem^B$ which preserve the ordering of the
  boundary components, and $\mathrm{PQC}^B_0(\riem^B)$ be the subspace
  of these which are isotopic to the identity relative to the
  boundary (that is, so that the isotopy fixes the boundary
  components pointwise).  Finally, let
  \[  \mathrm{PMod}^B(\riem^B)= \frac{\mathrm{PQC}^B(\riem^B)}
      {\mathrm{PQC}^B_0(\riem^B)}.  \]
 \end{definition}
We often abbreviate ``isotopic relative to the boundary'' by
``isotopy rel $\partial \riem^B$''. The ``P'' in the notation
stands for ``pure'',  which refers to the fact that the mapping
class group preserves the order of the boundary curves.
$\mathrm{PMod}^B(\riem^B)$ is often called the \textit{pure
(quasiconformal) mapping class group} or \textit{Teichm\"uller
modular group}.

 \begin{definition}
  For a Riemann surface $\riem^P$ with $n$ ordered punctures, let
  $\mathrm{PQC}^P(\riem^P)$ be the space of quasiconformal self-mappings
  of $\riem^P$ which preserve the punctures and their ordering, and
  $\mathrm{PQC}^P_0(\riem^P)$ be the subspace of these which are
  isotopic to the identity. Finally, let
  \[  \mathrm{PMod}^P(\riem^P)= \frac{\mathrm{PQC}^P(\riem^P)}
      {\mathrm{PQC}^P_0(\riem^P)}.  \]
 \end{definition}
We emphasize that throughout an isotopy the punctures must remain
fixed. This is automatic for a punctured surface but must be imposed as
an extra condition if one instead thinks of a surface with marked
points.

\begin{remark}
\label{rm:homo_iso} For a surface of finite topological type, the
Baer-Mangler-Epstein Theorem states that two orientation
preserving self-homeomorphisms are homotopic if and only if they
are isotopic (see \cite[Theorem 1.5.4]{Nag}).  With this in mind
we will use `isotopy' throughout this paper.  It is proved in
\cite{Earle_McMullen} that any homotopy can be replaced with an
isotopy such that the maps are uniformly quasiconformal, but we
will not need this fact.
\end{remark}

\begin{definition}  Let $\pmcgi[\riem^B]$
be the subgroup of $\mathrm{PMod}^B(\riem^B)$ consisting of
equivalence
 classes of quasiconformal mappings of $\riem^B$ whose representatives
 are the identity on $\partial \riem^B$.
\end{definition}

\begin{definition}
\label{DB} Let $\mathrm{DB}(\riem^B)$ be the subgroup of
$\pmcgi[\riem^B]$ generated by the equivalence classes of
mappings which are Dehn twists around curves that are isotopic to
boundary curves.
\end{definition}
Explicitly, let $\partial_i \riem^B$ be a boundary curve and
$\gamma_i$ be a curve isotopic to $\partial_i \riem^B$.  Let
$\ann_{\gamma_i}$ be the annular neighborhood bounded by
$\gamma_i$ and $\partial_i \riem^B$. $\mathrm{DB}(\riem^B)$
consists of maps which are equivalent in $\pmcgi[\riem^B]$ to
quasiconformal maps which are the identity on $\riem^B \backslash
\cup_i \ann_{\gamma_i}$.

\begin{definition}
\label{DI}
Let $\mathrm{DI}(\riem^B)$ be the subgroup of $\pmcgi[\riem^B]$
generated by Dehn twists around curves
$\gamma$ which are not isotopic to a boundary curve or a point.
These are the non-separating curves.
\end{definition}

As an aside, we note that the groups
$\mathrm{PMod}^P(\riem^P)$ and $\pmcgi[\riem^B]$ are isomorphic to their
analogues which are defined using homeomorphism or diffeomorphisms.

\begin{proposition}
\label{pr:mcg_generators} The mapping class group
$\pmcgi[\riem^B]$ is generated by Dehn  twists about
finitely many non-separating closed curves together with Dehn
twists about curves isotopic to the boundary components.
\end{proposition}

\begin{proposition}
\label{DB_central}
The subgroup $\mathrm{DB}(\riem^B)$ generated by boundary Dehn
twists is contained in the center of $\pmcgi[\riem^B]$ and is
isomorphic to $\mathbb{Z}^n$.
\end{proposition}
\begin{proof}
Dehn twists about disjoint curves commute because the twist
homeomorphisms can be taken to have disjoint support. The second
part is the content of \cite[Theorem 3.8]{ParisRolfsen}.
\end{proof}

\begin{corollary}
\label{MCG_DBDI}
$\pmcgi[\riem^B] / \mathrm{DB}(\riem^B) \simeq
\mathrm{DI}(\riem^B)$
\end{corollary}

\end{subsection}

\begin{subsection}{Teichm\"uller spaces}
\label{Teichmuller}

We define the ordinary Teichm\"uller spaces (i.e. without
riggings) of punctured and bordered surfaces and describe their
complex structure, as well as discuss the mapping class groups.
Since this material is standard we only provide a sketch, and
refer the reader to \cite{Lehto} or \cite{Nag} for details.

The Teichm\"uller space of a bordered Riemann surface is defined
as follows.  Let $\riem^B$ be a fixed base surface, which
establishes the genus and number of boundary components.  Consider
the set of triples $(\riem^B,f_1,\riem[1]^B)$ where $\riem[1]^B$
is a Riemann surface, and $f_1:\riem^B \rightarrow \riem[1]^B$ is
a quasiconformal mapping. We say that \[ (\riem^B,f_1,\riem[1]^B)
\sim_T (\riem^B,f_2,\riem[2]^B) \] if there exists a
biholomorphism $\sigma: \riem[1]^B \rightarrow \riem[2]^B$ such
that $f_2^{-1} \circ \sigma \circ f_1$ is isotopic to the identity
`rel $\partial \riem^B$'.  Recall that the term `rel $\partial
\riem^B$' means that the isotopy is constant on $\partial
\riem^B$; in particular it is the identity there.

\begin{remark}
If we impose an ordering of the boundary components of $\riem^B$ then
a map $f_1:\riem^B \rightarrow \riem[1]^B$ induces an ordering on
the boundary components of $\riem[1]^B$. In the definition of the
equivalence relation the `isotopy rel boundary' condition implies
that $\sigma$ automatically preserves the ordering.
\end{remark}

\begin{definition}
 The Teichm\"uller space of a bordered Riemann surface $\riem^B$
 is
 \[  T^B(\riem^B) \cong  \{ (\riem^B,f_1,\riem[1]^B) \}/\sim_T.  \]
\end{definition}
Taking the quotient by the weaker equivalence relation that $f_2^{-1}
\circ \sigma \circ f_1$ is isotopic to the identity (not necessarily
rel boundary), produces the \textit{reduced} Teichm\"uller
space $T_\#^B(\riem^B)$.

The case of punctured surfaces is similar. Let $\riem^P$ be the
punctured base surface. We say \[ (\riem^P,f_1,\riem[1]^P) \sim_T
(\riem^P,f_2, \riem[2]^P)\] if and only if there exists a
biholomorphism $\sigma : \riem[1]^P \to \riem[2]^P$ such that
$f_2^{-1} \circ \sigma \circ f_1$ is isotopic to the identity.
Note that the punctures are necessarily fixed throughout a
isotopy.
\begin{definition}
The Teichm\"uller space $T^P(\riem^P)$ of punctured Riemann
surfaces is
\[  T^P(\riem^P) \cong  \{ (\riem^P,f_1,\riem[1]^P) \}/\sim_T.  \]
\end{definition}

These two definitions are special cases of a more general
definition but we will not discuss this here.

The \textit{Beltrami equation} is the partial differential equation
$$
\frac{\partial w}{\partial \bar{z}} = \mu
\frac{\partial w}{\partial \bar{z}}.
$$
The following two theorems are crucial.
\begin{theorem}
\label{th:Beltrami_unique}
For any $\mu \in L^{\infty}(\Bbb{C})$ with $||\mu||_{\infty} <1$, there exists a unique solution to the Beltrami equation fixing $0,1$ and $\infty$.
This normalized quasiconformal map will be denoted $w^{\mu}$.
\end{theorem}
\begin{theorem}
\label{th:Beltrami_holo}
For every fixed $z \in \Bbb{C}$, the map $\mu \mapsto w^{\mu}(z)$ is
holomorphic. In particular, if $\mu$ depends on a parameter $t$
holomorphically, then $t \mapsto w^{\mu}(z)$ is holomorphic.
\end{theorem}

Let $X$ be a Riemann surface with punctures or boundary. We
temporarily denote its Teichm\"uller space by $T(X)$ and its mapping
class group by $\mathrm{PMod}(X)$. We have dropped the superscript
``B'' or ``P'' as the following considerations apply in both cases.

Let $L^{\infty}_{(-1,1)}(X)_1$ be the unit ball in the complex
Banach space of differentials of type $(-1,1)$ on $X$, which as a
linear space possesses a complex structure. Elements $\mu \,
d\bar{z}/dz \in L^{\infty}_{(-1,1)}(X)_1$ are called
\textit{Beltrami differentials}. If $f: X  \to X_1$ is
quasiconformal then, in terms of a local parameter $z$,
$$\mu(f) = \frac{\partial f}{\partial \bar{z}} / \frac{\partial f}{\partial z}$$
is called the \textit{complex dilation} of $f$.   The existence
and uniqueness of solutions  to the Beltrami equation
guarantees a well-defined association of an element
$[\riem^B,f_1,\riem[1]^B]$ to each element of
$L^{\infty}_{(-1,1)}(X)_1$. This association is called the
\textit{fundamental projection} and is denoted by
\[  \Phi : L^{\infty}_{(-1,1)}(X)_1 \longrightarrow T(X).  \]
The Teichm\"uller space $T(X)$ possesses a natural complex
structure, and this complex structure has the following crucial
properties.
\begin{theorem} \label{th:fundamentalprojectionholomorphic}
 The fundamental projection $\Phi : L^{\infty}_{(-1,1)}(X)_1
 \longrightarrow T(X)$ is holomorphic.
\end{theorem}
\begin{theorem} \label{th:localsectionsexistence}
 The fundamental projection possesses local holomorphic sections;
 i.e. for any point $p\in T(X)$ there is a holomorphic map
 $\sigma:U \rightarrow L^{\infty}_{(-1,1)}(X)_1$ on a
 neighborhood $U$ of $p$ such that
 $\Phi \circ \sigma$ is the identity.
\end{theorem}

Next, we define the `Teichm\"uller distance', a metric on
$T(X)$.  Any topological statements about the various Teichm\"uller
spaces in this paper
refer to the unique topology compatible with this distance.

For any quasiconformal mapping $f$ defined on a Riemann
surface $X$, we define its
`maximal dilatation' $K_f$ by
$$  || \mu(f)||_\infty =\frac{K_f-1}{K_f+1}  $$
where $\mu(f)$ is the complex dilatation of $\mu$.
\begin{definition} \label{de:Teichmullerdistance}
 The Teichm\"uller distance $\tau$ between two elements $[X,f,X_1]$ and
 $[X,g,X_2]$ of $T(X)$ is given by
 $$ \tau\left(  [X,f,X_1] ,[X,g,X_2] \right) = \frac{1}{2}\inf \{ \log{K_{g \circ f^{-1}}} \}$$
 where  the infimum is taken over all representatives $f$ and $g$ of the
 equivalence classes $[X,f,X_1]$ and $[X,g,X_2]$.
\end{definition}

A full exposition of the preceding material can be found in
\cite[Chapter 3]{Nag} or \cite[ Chapter V]{Lehto}.

 A quasiconformal map $h:X \to X$ induces an bijection
$h_* : T(X) \longrightarrow T(X)$ defined by  $[X,f,X_1] \mapsto
[X,f \circ h, X_1]$. For an element $[\rho] \in \mathrm{PMod}(X)$
we denote its corresponding action by $[\rho] \cdot [X,f,X_1] =
[X,f \circ \rho, X_1]$. (It is actually an anti-action but there
is no need to dwell upon this fact.) It is not hard to see that
this action is well defined and the quotient of $T(X)$ by this
action of the mapping class group is isomorphic to the Riemann
moduli space. The following Lemma is a deeper result (see for
example \cite[page 225]{Nag}).
\begin{lemma}
\label{Modaction_holo} The group $\mathrm{PMod}(X)$ acts as a
group of biholomorphisms on $T(X)$. That is, $\rho_* : T(X) \to
T(X)$ is a biholomorphism for each $\rho \in \mathrm{PMod}(X)$.
\end{lemma}

Since there are many definitions of proper discontinuity to choose
from,  we include the following definition for definiteness.
\begin{definition}
\label{de:propdisc}
The action of a group $G$ on a topological space $S$ is called \textit{properly discontinuous}
if for any $s \in S$ there exists a neighborhood $V_s$ of $s$ such that
$(g \cdot V_s) \cap V_s = \emptyset$ except for finitely many $g \in G$.
\end{definition}

The next result can be found in \cite[page 152]{Nag}.
\begin{lemma}
\label{ModactionTP}
Let $\riem^P$ be a punctured Riemann surface of finite type.
The action of the group $\mathrm{PMod}^P(\riem^P)$ on $T^P(\riem^P)$
is properly discontinuous.
\end{lemma}

The behavior of the action of $\pmcgi[\riem^B]$ and its subgroups
on $T^B(\riem^B)$ is an important part of this paper. The relation
to $\mathrm{PMod}^P(\riem^P)$ will also be discussed and utilized.
For the structure of certain quotient spaces we will need the following
(see \cite[Proposition 5.3]{Wells} or \cite[page 160]{Nag}).

\begin{proposition}
\label{complex_quotient}
Let $M$ be a complex manifold. Let $G$ be a group acting properly
discontinuously and fixed-point freely by biholomorphisms on $M$.  Then
$M/G$ is a Hausdorff topological space which can be given a unique
complex structure, so that the projection mapping $p: M \to M/G$ is
holomorphic. Moreover, $p$ possess local holomorphic sections.
\end{proposition}

\begin{remark}
\textbf{} Proposition \ref{complex_quotient} holds in the case
that $M$ is infinite-dimensional. Uniqueness relies on the fact that a homeomorphism is
holomorphic if it is holomorphic on all finite-dimensional affine
subspaces (see \cite[page 87]{Nag}).
\end{remark}
\end{subsection}


\begin{subsection}{Holomorphic motions and the $\lambda$-lemma}

The $\lambda$-lemma and its extension are discussed.  This result
will be used in a fundamental way in this section to produce
quasiconformal maps with specified boundary values. The material
in this section is taken from \cite{AM01} and \cite{BR86}.
Originally the $\lambda$-lemma is due to Ma\~n\'e, Sad and
Sullivan \cite{MSS} where it was used in the context of complex
dynamics.

\begin{remark}
The $\lambda$-lemma was used in a different way in
\cite{Radnell_thesis} for proving holomorphicity of the sewing
operation. Similar ideas will be used in Section
\ref{se:compat_thesis} to demonstrate the compatibility of the complex
structure on the rigged moduli space with those given in
\cite{Huang} and \cite{Radnell_thesis}.
\end{remark}

Let $\disk$ be the open unit disk in $\Bbb{C}$.
\begin{definition}
\label{holo_motion} Let $A$ be a subset of $\hat{\Bbb{C}}$. A
holomorphic motion of $A$ is a map $f:\disk \times A \rightarrow
\hat{\Bbb{C}}$ such that:
\begin{enumerate}
\item for any fixed $z \in A$, the map $t \mapsto f(t,z)$ is
holomorphic on $\disk$, \item for any fixed $t \in \disk$, the map
$z \mapsto f(t,z)$ is an injection, and \item the mapping $f(0,z)$
is the identity on $A$.
\end{enumerate}
\end{definition}
Since $t$ is a kind of deformation parameter we
often use the notation $f_t(z)$ for $f(t,z)$.
Also, as $f_0$ is the identity, we think of $f_t(z)$ as a holomorphic
perturbation of the identity.
The following theorem is the $\lambda$-lemma of \cite{MSS}.
It says that any holomorphic perturbation of the identity
must be a quasiconformal map.

\begin{theorem}[$\lambda$-lemma]
\label{lambda-lemma} If $f$ is a holomorphic motion as above then
$f$ has an extension to $F:\disk \times \overline{A} \to
\hat{\Bbb{C}}$ such that:
\begin{enumerate}
\item $F$ is a holomorphic motion of $\overline{A}$,
\item each $F_t(\cdot) : \overline{A} \to \hat{\Bbb{C}}$ is
quasiconformal, and
\item $F$ is jointly continuous in $(t,z)$.
\end{enumerate}
\end{theorem}

In fact the holomorphic motion extends to the whole plane.
This was originally proved by S{\l}odkowski in \cite{Slodkowski}
although other proofs now exist.

\begin{theorem}[Extended $\lambda$-lemma]
\label{th:exlambda} If $f$ is a holomorphic motion as above then
$f$ has an extension to $F:\disk \times \hat{\mathbb C}
\rightarrow \hat{\mathbb C}$ such that:
\begin{enumerate}
\item $F$ is a holomorphic motion of $\hat{\mathbb C}$,
\item each $F_t:\hat{\mathbb C} \rightarrow \hat{\mathbb C}$ is
 quasiconformal with dilatation not exceeding $(1+|t|)/(1-|t|)$, and
\item $F$ is jointly continuous in $(t,z)$.
\end{enumerate}
\end{theorem}

The Beltrami differential of the
quasiconformal extension $\mu(F_t)$ is holomorphic in $t$.  See for
example \cite[Theorem 2]{BR86}.
\begin{theorem}
\label{th:holomotion_dilation}
Let $f$ be a holomorphic motion of a set $A$ with non-empty interior
$A^0$. Then in $A^0$, the Beltrami coefficient $\mu(t,z)$ of $f(t,z)$
is a holomorphic function of $t \in \disk$.  That is, the map $ \disk
\to L^{\infty}(A^0)_1 $ given by $t \mapsto \mu(t,z)$ is holomorphic.
\end{theorem}
Note that in particular this theorem applies to $F_t(z)$ in
the extended $\lambda$-lemma and so $\mu(F_t) \in L^{\infty}(\Chat)_1$ is holomorphic in $t$.

\end{subsection}


\begin{subsection}{Quasisymmetric maps}

We briefly review some standard definitions and adjust them to our
purposes. Useful facts about extensions of quasiconformal maps
are stated.
Let $\overline{\mathbb R}$ denote the extended real line
${\mathbb R} \cup \infty$.

\begin{definition} \label{quasisymmetricline}
 An (orientation preserving)
 homeomorphism
$$h:\overline{\mathbb R} \rightarrow
 \overline{\mathbb R}
$$ is $k$-quasisymmetric if there exists a
 constant $k$ such that
 \[  \frac{1}{k} \leq \frac{h(x+t)-h(x)}{h(x)-h(x-t)} \leq k  \]
 for all $x,\,t \in \overline{\mathbb R}$.  If $h$ is
 quasisymmetric for some unspecified $k$ it is simply called
 quasisymmetric.
\end{definition}

We find it more convenient to work on $S^1$ than on
$\overline{\mathbb R}$.  It is also necessary to speak of
quasisymmetry of a mapping on a closed boundary curve of a Riemann
surface. The map $T(z) = i(1+z)/(1-z)$ sends the unit circle to
$\overline{\Bbb{R}}$ with $T(1) = \infty$.
\begin{definition} \label{quasisymmetriccircle}
 Let $h: S^1 \rightarrow S^1$ be a homeomorphism.
 \begin{enumerate}
 \item Let $e^{i\theta}$ be chosen so that $e^{i\theta}h(1)=1$.
Then we say that $h$ is quasisymmetric
 if $T \circ e^{i\theta} h \circ T^{-1}$ is quasisymmetric according to
 Definition \ref{quasisymmetricline}.
 \item Let $C$ be a connected component of the border of a Riemann
 surface, and $h$ a homeomorphism of $C$ into $S^1$.  We say that
 $h$ is quasisymmetric if, for any biholomorphism $F: \ann_C
 \rightarrow \ann^1_r$ from a neighborhood $\ann_C$ of $C$ into a
 standard annulus with outer radius $1$, $h \circ F^{-1}$ is
 quasisymmetric on $S^1$ in the sense of part one.
 \end{enumerate}
\end{definition}

Three observations should be made at this point.

\begin{remark}[Regularity of the inner boundary curve]
Given $F$, we can always restrict to a smaller neighborhood such
that the inside boundary of $\ann_C$ is an analytic curve.
\end{remark}

\begin{remark}[Independence of the choice of $F$]  The boundary values
 of a biholomorphism of a sufficiently nice domain are quasisymmetric.
 In particular, if $H: \ann^1_{C_1} \rightarrow \ann^1_{C_2}$ is a
 biholomorphic map between doubly connected domains bounded by $S^1$
 and analytic curves $C_i$, then $H$ extends to a quasisymmetric map
 on $S^1$.  Also the composition of two quasisymmetric functions is
 quasisymmetric \cite[II.7]{LV}.  Thus if part two of Definition
 \ref{quasisymmetriccircle} holds for one biholomorphism $F_1$, it
 holds for any biholomorphism $F_2$ of an annular neighborhood,
 since $h \circ F_2^{-1}=h \circ F_1^{-1} \circ F_1 \circ F_2^{-1}$.
In fact, this is true in greater generality: it holds if $H$ is
 quasiconformal (see Theorem \ref{qsisboundaryqc} ahead), and further
 if $C_i$ are quasicircles.
\end{remark}
\begin{remark}[On $k$-quasisymmetry] These definitions cannot be
 refined to $k$-quasisymmetry in a canonical way.
\end{remark}

The following theorem explains the importance of quasisymmetric
mappings in Teichm\"uller theory (see \cite[II.7]{LV}).
\begin{theorem} \label{basicqcextension}
 A homeomorphism $h:\overline{\mathbb R} \rightarrow
 \overline{\mathbb R}$ is quasisymmetric if and only if there
 exists a quasiconformal map of the upper half plane with boundary
 values $h$.
\end{theorem}
Since quasiconformality is preserved under biholomorphisms, a map
$h: S^1 \rightarrow S^1$ is quasisymmetric if and only if there is
a quasiconformal map from the unit disk to itself with boundary
values $h$.

We require a local version of this statement, involving only a
doubly connected neighborhood of a closed boundary curve. Towards
this end we collect some more results on extensions of
quasiconformal mappings.  We will also require a result on
continuation of quasiconformal maps across certain sets of measure
zero. The following results are taken directly from Theorems 8.1,
8.2 and 8.3 in \cite[II.8]{LV}.

\begin{definition}
A curve $\gamma \subset \Bbb{C}$  in called a \textit{quasicircle}
if it is the image of a circle under a quasiconformal mapping of the plane.
A \textit{quasiarc} is a subarc of a quasicircle.
\end{definition}
Note that in \cite{LV} quasiarcs are called quasiconformal curves.

\begin{theorem}
\label{LV_ext1} Let $w_0 : G \rightarrow G'$ be a quasiconformal
mapping and $F$ a compact subset of the domain $G$.  Then
$w_0|_{F}$ extends to a quasiconformal mapping of the whole plane.
\end{theorem}

\begin{theorem}
\label{LV_ext2} Let $G$ and $G'$ be two domains with free boundary
curves $C$ and $C'$. Let $w : G \longrightarrow G'$ be a
quasiconformal mapping such that $w(C) = C'$. If $C$ and $C'$ are
quasicircles then $w$ can be extended to a quasiconformal mapping of
any domain $G_1$ containing $G \cup C$.
\end{theorem}

Combining the above two theorems we obtain the following.
\begin{theorem}
\label{LV_ext3} Let $G$ and $G'$ be two $n$-tuply connected
domains whose boundary curves are quasicircles. Then every
quasiconformal mapping $w : G \longrightarrow G'$  can be extended
to a quasiconformal mapping of the whole plane.
\end{theorem}

The next theorem states that quasiarcs are removable for
quasiconformal mappings. See for example \cite{Mori},
\cite{Strebel} or \cite[V.3]{LV}.

\begin{theorem}
\label{th:qc_remove} Let $G$ be an open subset of $\Bbb{C}$ and
let $E$ be a closed subset of $G$ such that the two-dimensional
measure of $E$ is zero. If $w$ is a homeomorphism of $G$ that is
$K$-quasiconformal on $G\setminus E$, then $w$ is $K$-quasiconformal on
$G$. In particular this holds when $E$ is a quasiarc.
\end{theorem}

 We now return to the problem of giving a version of Theorem
 \ref{basicqcextension} for boundary curves of a Riemann
 surface. Consider any quasiconformal map $f:\ann^1_{C_1} \rightarrow
 \ann^1_{C_2}$ where $C_i$ are Jordan curves enclosed by
 $S^1$. Theorems \ref{LV_ext1} and \ref{LV_ext2} guarantee that for
 any $r$ such that $\ann^1_r \subset \ann^1_{C_1}$, there exists an
 extension of $f$ to the disk, $\disk$, which agrees with $f$ on
 $\ann^1_r$. So we have the following theorem.

\begin{theorem} \label{qsisboundaryqc} \
 \begin{enumerate}
 \item A map $h: S^1 \rightarrow S^1$ is quasisymmetric if and only if
 it is the restriction of some quasiconformal map $f: \ann^1_{C_1}
 \rightarrow \ann^1_{C_2}$ for Jordan curves $C_1$ and $C_2$ enclosed
 by $S^1$.
 \item Let $C$ be a connected component of the border of a Riemann
 surface and $h:C \rightarrow S^1$ be a homeomorphism.  The map $h$ is
 quasisymmetric if and only if it is the restriction of some
 quasiconformal map $f: \ann_C \rightarrow
 \ann^1_{C_1}$ on some doubly connected neighborhood $\ann_C$ of
 $C$.
 \end{enumerate}
\end{theorem}

\begin{remark}
 Definition \ref{quasisymmetriccircle} and Theorem \ref{qsisboundaryqc}
 can be extended in the obvious way to mappings between connected
 components of a bordered Riemann surface.
\end{remark}

\end{subsection}
\end{section}


\begin{section}{Quasisymmetric sewing}
\label{sewing}

As discussed in the introduction, the operation of sewing Riemann
surfaces along analytically parametrized boundary components is a
fundamental operation in conformal field theory.  In this section
it is shown how to sew Riemann surfaces where the identification
of the boundary components is by quasisymmetric maps.  For the
case of disk this is a well-known construction in geometric
function theory called  \textit{conformal welding}
\cite[III.1.4]{Lehto}, \cite{LVsewingpaper}, \cite{Pfluger}.
Sewing the boundaries of a strip to produce annuli was
investigated by Oikawa \cite{Oikawa}. The case of higher-genus
surfaces is no more difficult but, at least in the context of
conformal field theory, it has not been discussed in the
literature.

A second fundamental use of sewing is to produce punctured surfaces by
sewing caps onto bordered surfaces. Being able to do this with
quasisymmetric maps enables us to relate the puncture and border models of
rigged Teichm\"uller space.


\begin{subsection}{The sewing operation}
\label{sewingoperation}
Let $\disk^* = \Chat \setminus \disk$ be the upper-hemisphere.
The following theorem (see \cite[III.1.4]{Lehto}) describes the
classical conformal welding of disks.

\begin{theorem}
\label{th:welding} If $h : S^1 \longrightarrow S^1$ is
quasisymmetric then there exists conformal maps $F$ and $G$ from
$\disk$ and $\disk^*$ into complementary Jordan domains $\Omega$
and $\Omega^*$ of $\Chat$ such that $G^{-1} \circ F|_{S^1} = h$.
Moreover, the Jordan curve separating $\Omega$ and $\Omega^*$ is a
quasicircle.
\end{theorem}

We now describe the sewing of arbitrary Riemann surfaces using the
conformal welding idea. Let $\riem[1]^B$ and $\riem[2]^B$ be
bordered Riemann surfaces of type  $(g_1,n^-_1,n^+_1)$
and $(g_2,n^-_2,n^+_2)$ respectively where $n^+_1 > 0 $ and
$n^-_2>0$. Let $C_1$ be an outgoing boundary component of
$\riem[1]^B$ and $C_2$ be an incoming boundary component of
$\riem[2]^B$. Let $\psi_1$ and $\psi_2$ be quasisymmetric
parametrizations of $C_1$ and $C_2$ (that is, quasisymmetric maps
$\psi_i: C_i \to S^1$, for $i = 1,2$, in the sense of Definition
\ref{quasisymmetriccircle}). Define $J : \Chat \to \Chat$ by $J(z)
= 1/z$ and note that $\psi_2^{-1} \circ J \circ \psi_1 : C_1 \to
C_2$ is an orientation reversing map.

\begin{figure}[tp]
\begin{center}
\input{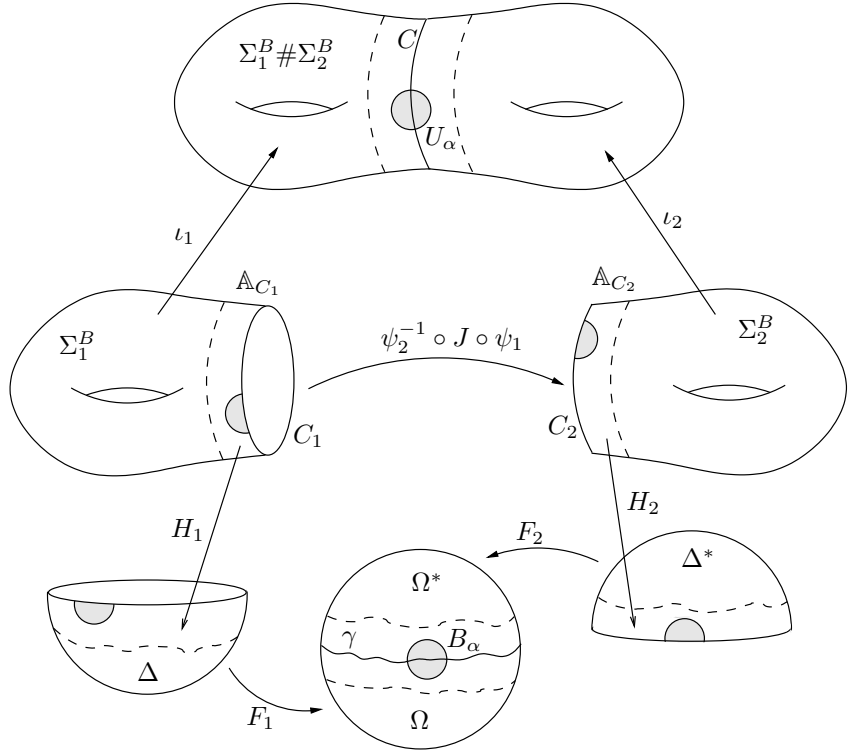}
\caption{Sewing using quasisymmetric boundary identification.}
\label{sewing_figure}
\end{center}
\end{figure}

\begin{remark}
\label{re:CFTorientation}
In conformal field theory it is
customary to include the information of `incoming/outgoing' (that is,
choice of sign) in the orientation of the parametrization. For our
purposes it is easier to work solely with positively oriented
quasisymmetric maps and record the orientation separately (see
Remark \ref{re:inout}). This is the view taken in \cite{Huang} and is
the most natural approach when working in the puncture picture of the
rigged moduli space.  We could recover the standard picture by
left-composing all the parametrizations, of the incoming boundary,
with $J$.
\end{remark}

Let $\riem[1]^B \# \riem[2]^B = \riem[1]^B \sqcup \riem[2]^B /
\sim$ where $x \sim y$ if and only if $x \in C_1$, $y \in C_2$ and
$(\psi_2^{-1} \circ J \circ \psi_1)(x) = y$. Let $\iota_i$ be the
inclusion maps $\riem[i]^B \longrightarrow \riem[1]^B \#
\riem[2]^B$. Both $C_1$ and $C_2$ map to a common curve on
$\riem[1]^B \# \riem[2]^B = \riem[1]^B \sqcup \riem[2]^B / \sim$
which we denote by $C$ and call the \textit{seam}.

\begin{remark}
\label{Ahlfors_sewing}
There is a natural way to make $\riem[1]^B \# \riem[2]^B$ into a
topological space. If $\psi_1$ and $\psi_2$ are analytic
parametrizations then $\riem[1]^B \# \riem[2]^B$ becomes a Riemann surface
in a standard way using $\psi_1$ and $\psi_2$ to produce charts on the
seam, $C$. See for example Ahlfors and Sario \cite[Section II.3D]{Ahlfors_Sario}.
\end{remark}

We now describe how to put a complex structure on the sewn
surface. Consulting Figure \ref{sewing_figure} may be helpful. Let
$\ann_{C_1}$ and $\ann_{C_2}$ be annular neighborhoods of $C_1$
and $C_2$ and choose biholomorphic maps $ H_1 : \ann_{C_1}
\longrightarrow \disk $ and $H_2 : \ann_{C_2} \longrightarrow
\disk^* $ such that $H_i(C_i) = S^1$.  Note that the images of
$H_1$ and $H_2$ are annular neighborhoods of $S^1$.  With the
orientation on $C_i$ induced from $\riem[i]^B$, $H_1|_{C_1} : C_1
\to S^1$ is orientation preserving while $H_2|_{C_2} : C_2 \to
S^1$ is orientation reversing.

The map
\begin{equation}
\label{h}
h = H_2 \circ \psi_2^{-1} \circ J \circ \psi_1 \circ H_1^{-1}|_{S^1} : S^1
\longrightarrow S^1
\end{equation}
is orientation preserving and is quasisymmetric since it is the
composition of quasisymmetric maps. By Theorem \ref{th:welding}
there exist conformal maps $F_1$ and $F_2$ from $\disk$ and
$\disk^*$ into complementary Jordan domains in $\Chat$ such that
$h = F_2^{-1} \circ F_1|_{S^1}$. Let $\gamma$ be the quasicircle
separating the Jordan domains $\Omega$ and $\Omega^*$.

The charts on $\riem[1]^B \# \riem[2]^B$ will now be described. On
$\iota_1(\riem[1]^B)$ or $\iota_2(\riem[2]^B)$ we take the original
(interior) charts on $\riem[1]^B$ and $\riem[2]^B$.  To be precise we
really have to compose with the inclusions $\iota_i$.  On the join we
must be more careful. We consider all sets, $B_{\alpha}$, of the
following form.  Let $B_{\alpha}$ be a simply connected domain in
$\Chat$ such that $B_{\alpha} \cap \gamma \neq \emptyset$ and
$B_{\alpha} \subset F_1(H_1(\ann_{C_1})) \cup
F_2(H_2(\ann_{C_2}))$. Let $B_{\alpha}^+ = B_{\alpha} \cap
\overline{\Omega}$ and $B_{\alpha}^- = B_{\alpha} \cap
\overline{\Omega^*}$.  We can use these half-balls to define a chart in
a neighborhood of a point in $C$. Let $A = \iota_1(\ann_{C_1}) \cup
\iota_2(\ann_{C_2})$, and note that it is an annular neighborhood of
$C$. For $i=1,2$, define $\zeta_i : A \to \Chat$ by
\begin{equation}
\label{zeta_i}
\zeta_i = F_i\circ H_i \circ \iota_i^{-1},
\end{equation}
and let $ U_{\alpha}^+ = \zeta_1^{-1}(B_{\alpha}^+) $ and $
U_{\alpha}^- = \zeta_2^{-1}(B_{\alpha}^-)$.  Let $U_{\alpha} =
U_{\alpha}^+ \cup U_{\alpha}^-$ and define $\zeta_{\alpha} :
U_{\alpha} \longrightarrow B_{\alpha}$ by
\begin{equation}
\label{zeta_alpha}
\zeta_{\alpha}(p) =
\begin{cases}
\zeta_1(p) & \text{if } p \in   U_{\alpha}^+ \\
\zeta_2(p) & \text{if } p \in   U_{\alpha}^- .
\end{cases}
\end{equation}
We take all such $U_\alpha$ together
with the original (interior) open sets on
$\riem[1]^B$ and $\riem[2]^B$ to form a basis of open sets and hence a
topology on $\riem[1]^B \# \riem[2]^B$.  In this topology we have:
\begin{lemma}
\label{le:zeta_homeo}
The map $\zeta_{\alpha} :  U_{\alpha} \longrightarrow B_{\alpha}$
is a homeomorphism.
\end{lemma}
\begin{proof}
Since $\zeta_{\alpha}$ is constructed from two homeomorphisms we need
only check that it is well defined on $U_{\alpha}^+ \cap U_{\alpha}^-
\subset C$. Let $w \in U_{\alpha}^+ \cap U_{\alpha}^-$ and let
$x$ and $y$ be points in $C_1$ and $C_2$ respectively such that
$\iota_1(x) = \iota_2(y) = w$.
This implies $\psi_1(x) = J \circ \psi_2(y)$.
We need to show that $\zeta_1(w) = \zeta_2(w)$.

By definition of $\zeta_{\alpha}$ we must show that
$(F_1\circ H_1) (x) =  (F_2\circ H_2)(y)$.
The conformal welding maps $F_1$ and $F_2$ have the property that
$F_2^{-1} \circ F_1(z) = h(z)$ where
$h$ is defined in \eqref{h}.
Therefore
\begin{align*}
(F_1\circ H_1) (x) & = ((F_2\circ h) \circ H_1)(x) \\
& =
(F_2 \circ ( H_2 \circ \psi_2^{-1}\circ J \circ \psi_1 \circ H_1^{-1}) \circ H_1)(x)
\\
&= ((F_2 \circ H_2) \circ  (\psi_2^{-1} \circ J \circ \psi_1)) (x) \\
&= (F_2 \circ H_2)(y)
\end{align*}
as required.
\end{proof}

\begin{theorem}
\label{th:complex_structure}
The charts $(U_{\alpha}, \zeta_{\alpha})$ together with the original
(interior) charts from $\riem[1]^B$ and $\riem[2]^B$ give $\riem[1]^B
\# \riem[2]^B$ a complex manifold structure.  That is, $\riem[1]^B \#
\riem[2]^B$ with these charts is a Riemann surface.
\end{theorem}
\begin{proof}
The transition function corresponding to a chart $(U_{\alpha},
\zeta_{\alpha})$, and a chart from $\riem[1]^B$ or $\riem[2]^B$, is
holomorphic because it is a composition of functions that are holomorphic
with respect to the complex structures on $\riem[1]^B$ or $\riem[2]^B$.

Let $(U_{\alpha}, \zeta_{\alpha})$ and $(U_{\beta},
\zeta_{\beta})$ be two charts with $U_{\alpha} \cap U_{\beta} \neq
\emptyset$.  From equations \eqref{zeta_i} and
\eqref{zeta_alpha} we see that the $\zeta_{\alpha}$ are defined
using the globally defined maps $F_i$, $H_i$ and $\iota_i$.
Moreover, the transition function $\zeta_{\alpha} \circ
\zeta_{\beta}^{-1}$ is the identity and is thus holomorphic.
\end{proof}

\begin{remark}
If $\psi_1$ and $\psi_2$ are analytic then we can choose $H_1 =
\psi_1$ and $H_2 = J \circ \psi_2$. In this case $h = \text{id}$, $F_1 =
\text{id}$ and $F_2= \text{id}$. So our more general sewing procedure
reduces to the standard one outlined in Remark \ref{Ahlfors_sewing}.
\end{remark}

\begin{theorem}
The complex structure on $\riem[1]^B \# \riem[2]^B$ defined
in Theorem \ref{th:complex_structure}
is the unique
complex structure which is compatible with the original complex structures
on $\riem[1]^B$ and $\riem[2]^B$.
\end{theorem}
\begin{proof}
Let $\{(V_{\beta}, \xi_{\beta})\}$ be an analytic atlas for
$\riem[1]^B \# \riem[2]^B$ that is compatible with the complex
structures on $\riem[1]^B$ and $\riem[2]^B$.  The compatibility means that
if $V_{\beta} \subset \iota_i(\riem[i]^B)$ then $\xi_{\beta}\circ
\iota_i $ is a holomorphic function from $\riem[i]^B$ to $\Bbb{C}$.

The case that needs attention is for the transition functions for
the charts $(U_{\alpha}, \zeta_{\alpha})$ and $(V_{\beta},
\xi_{\beta})$ where $U_{\alpha} \cap V_{\beta} \cap C \neq
\emptyset$. The compatibility of $\zeta_{\alpha}$ and
$\xi_{\beta}$ with the complex structures on $\riem[1]^B$ and
$\riem[2]^B$ means that at points not in $C$, the holomorphicity
of the transition function $\xi_{\beta} \circ \zeta{_\alpha}^{-1}$
is immediate.  On the other hand let $x \in C \cap (U_{\alpha}
\cap V_{\beta})$ and let $p = \zeta_{\alpha}(x)$. We need to show
that $\xi_{\beta} \circ \zeta{_\alpha}^{-1}$ is holomorphic at
$p$. Let $N_{\alpha} = \zeta_{\alpha}(U_{\alpha} \cap V_{\beta})$
and $N_{\beta} = \xi_{\beta} (U_{\alpha} \cap V_{\beta})$.  By
Lemma \ref{le:zeta_homeo} the transition map $ \xi_{\beta} \circ
\zeta_{\alpha}^{-1} : N_{\alpha} \rightarrow N_{\beta} $ is a
homeomorphism and its restrictions to $N_{\alpha}^+ = N_{\alpha}
\cap \Omega$ and $N_{\alpha}^- = N_{\alpha} \cap \Omega^*$ are
holomorphic. By Theorem \ref{th:welding} we know that the boundary
$\partial N_{\alpha}^+ \cap \Omega$ is a quasiarc.  Theorem
\ref{th:qc_remove} therefore implies that $\zeta_{\alpha}^{-1}
\circ \xi_{\beta}$ is in fact holomorphic on all of $N_{\alpha}$.
\end{proof}

\begin{remark}
In the case of welding to produce annuli, an analogous result is
proved in \cite[Lemma 2 and Theorem 1]{Oikawa}. A key
ingredient in that case is also the removability of quasiarcs for
quasiconformal maps.
\end{remark}

\end{subsection}


\begin{subsection}{Sewing on caps}
\label{capsewing}
We describe in detail the sewing of caps onto a surface with boundary
to produce a punctured surface.  The general setup and notation will
be used throughout the paper.

The punctured disk $\overline{\disk}_0 = \{z \in \Bbb{C} \st 0 <
|z| \leq 1 \}$ will be considered as a bordered Riemann surface
whose boundary is parametrized by the identity map. Given
$\riem^B$ of type $(g,n^-,n^+)$ we choose a collection of
`boundary trivializations' $\vc{\tau}=(\tau_1,\ldots, \tau_{n})$;
each component is a quasiconformal map from a collared
neighborhood of a boundary curve to the annulus $\ann^1_r$ for
some $r<1$.  By Theorem \ref{qsisboundaryqc} we could equivalently
say $\tau_i : \partial_i \riem^B \to S^1$ is a quasisymmetric
boundary parametrization.

At each boundary curve $\partial_i \riem^B$ we sew in the punctured
disk $\overline{\disk}_0$ using $\text{id} \circ J \circ \tau_i$ as
the identification map as described in Section \ref{sewingoperation}.
We denote the simultaneous sewing by $\riem^B\#_\tau (\cdisk_0)^{n}$
and let $\riem^P=\riem^B\#_\tau (\cdisk_0)^{n}$ be the resultant
punctured surface. The images of the punctured disks in
$\riem^P$ will be called \textit{caps}.  Let $D_i$ denote
the $i$th cap,
$\mathbf{D} = D_1 \cup \ldots \cup D_{n}$ and $\mathbf{D}^0$ be the interior of $\mathbf{D}$.
Note that  $\riem^B = \riem^P \setminus \mathbf{D}^0$.
With some slight abuse of notation we
use $\partial_i \riem^B$ to denote the image of the $i$th boundary curve in $\riem^P$.

\end{subsection}


\end{section}

\begin{section}{Quasiconformal extensions}
\label{qs_extensions}

Here we collect some non-standard results on quasiconformal maps
and prove some new ones that we need. A key tool is the extended
$\lambda$-lemma.

Two of the important technical results of this section are the
related Lemmas \ref{correctingmap} and
\ref{strengthenedcorrectingmap}, which address the problem of
deforming quasiconformal maps to take specified boundary values.
Their immediate application is to the relation between  the
mapping class groups of a bordered surface and its corresponding
punctured surface. Moreover, these results are crucial in showing
the equivalence between the border and puncture models of rigged
Teichm\"uller space in Section \ref{ModuliandTeichmuller}.


\begin{subsection}{Holomorphic families and extension results}

The main idea here is that any quasisymmetric/quasiconformal map
can be embedded in a holomorphic motion. Combining this with the
extended $\lambda$-lemma, we are able to show that a
quasisymmetric mapping of the boundary of an annulus extends to
the entire annulus.

A version of the following Lemma can
be found in \cite[Example 1]{Pommerenke_Rodin}. Our inclusion of
the normalization is important in the application of the result.

\begin{lemma}
\label{le:holo_embedding} Any quasiconformal map $u : \Bbb{C}
\rightarrow \Bbb{C}$ can be embedded in a holomorphic motion. That
is, there exists a holomorphic motion $u^t(z)$ such that $u(z) =
u^{t_0}(z)$ for some $t_0 \in \disk$.  Furthermore, if $u(0)=0$,
we may take holomorphic motion to satisfy $u^t(0)=0$ for all $t$.
\end{lemma}
\begin{proof}
We proceed in two steps.  First we normalize $u$ and then find a
holomorphic family relating $u$ to the normalizing map. Second we
produce a holomorphic motion of the normalizing map.  Combining
these produces the required holomorphic motion of $u$.

Let $\sigma$ be the M\"obius transformation such that $\sigma
\circ u$ fixes $0$, $1$ and $\infty$. Actually since $u$ already
fixes $\infty$, $\sigma(z) = az + b$ for some $a,b \in {\mathbb
C}$.

The complex dilation $\mu(\sigma \circ u)$ is equal to $\mu(u)$
because $\sigma$ is conformal (but actually this is not important in
our case). By the definition of quasiconformality,
$||\mu(u)||_{\infty} = k $ for some $k<1$.  Choose $\ell$ such that $
k < \ell < 1$ and let $n=1/\ell$.

For $t \in \disk$, $\mu^t = nt\mu(u) $ defines a holomorphic
family of Beltrami coefficients. This follows since $||nt
\mu(u)||_{\infty} < |n||t||k| < 1$. The factor $n$ ensures that
$\mu^t = \mu(u)$ when $t = \ell$, which is in the interior of
$\disk$.

Let $w^t$ be the normalized solution to the Beltrami equation with
coefficient $\mu^t$ (see Theorem \ref{th:Beltrami_unique}).
The following properties hold:
\begin{enumerate}
\item for fixed $z$, $w^t(z)$ is holomorphic in $t$ (see Theorem \ref{th:Beltrami_holo}),
\item $w^0(z)= z$, and
\item $w^{\ell} = \sigma \circ u$.
\end{enumerate}

We now need to embed $\sigma^{-1}$ in a holomorphic motion. Let
$$
\rho^t(z) = \frac{(\ell-t)z + t\sigma^{-1}(z)}{\ell}
$$
and note that $\rho^t$ is also holomorphic in $z$. We claim that
$u^t(z) = (\rho^t \circ w^t)(z)$
is the desired holomorphic motion. To see this note that:
\begin{enumerate}
\item for fixed $z$, $u^t(z)$ is holomorphic in  $t$ because
$\rho^t(z)$ is holomorphic in $t$ and $z$,
\item $u^0(z) = z$, and
\item $u^{\ell} = \sigma^{-1} \circ w^{\ell} = u$.
\end{enumerate}

The final statement of the theorem follows from the fact that if
$u(0)=0$, then $\sigma(0)=0$.  Since $w_t$ is normalized so that
$w_t(0)=0$ for all $t$, we have $u_t(0)=0$ for all $t$.
\end{proof}

\begin{remark}
This proves that any quasicircle can be embedded in a holomorphic
motion of $S^1$.
\end{remark}

\begin{remark}
The next lemma (Lemma \ref{le:annulus_extension}) follows
independently from two different results in \cite{Tukia_Vaisala}.
The first result \cite[Lemma 3.3]{Tukia_Vaisala} is the much more
general one, and has a highly non-trivial proof.  The second
result \cite[Theorem 3.14]{Tukia_Vaisala} is a version of Lemma
\ref{le:annulus_extension} for arbitrary dimension. It has an
elementary proof, but there is a gap for $n=2$ (the case we are
interested in). The factorization of quasisymmetric maps $S^1 \to
S^1$ into quasisymmetries that are the identity on some arc of
$S^1$ must be proved. This factorization can be achieved using
standard approximation techniques.  We thank J. Heinonen for
pointing us to this literature and explaining how to fill the gap.

However, instead of following this approach, we give an elementary proof of Lemma \ref{le:annulus_extension} using completely different techniques (the extended $\lambda$-lemma).
\end{remark}

Choose $R>1$ and consider the standard annuli $\ann_1^R$. Recall
that $S_R$ is the circle of radius $R$ and $\overline{B(0,R)}$ is
the closed ball of radius $R$.
\begin{lemma}
\label{le:annulus_extension}
 Let $f : S_R \to S_R$ be a quasisymmetric mapping and let
 $\iota : S_1 \to S_1$ be the identity map.
 There exists a
 quasiconformal mapping $F: \ann_1^R \to \ann_1^R$
 extending $f_1$ and $\iota$. That is, $F|_{S_1} = \iota$ and
 $F|_{S_R} = f$. In fact $F$ can be extended to $\hat{{\Bbb{C}}}$.
\end{lemma}
\begin{proof}
The idea is to embed $f$ in a holomorphic motion and apply the
extended $\lambda$-lemma to the motion of $S_1\cup S_{R}$, with the motion
being the identity on $S_1$.  However this cannot be done directly
as the motion of $S_R$ may intersect $S_1$ and so the motion of
$S_1\cup S_{R}$ may not be injective.  We use compactness of the
motion and a rescaling to avoid this problem.

Let $u_f : \Bbb{C} \to \Bbb{C}$ be a quasiconformal extension of
$f$. We may assume that $u_f(0) = 0$ (this can be achieved by
composing $u_f$ with an appropriate linear transformation).

From Lemma \ref{le:holo_embedding} there exists a holomorphic
motion $u^t$ such that $u^{t_0} = u_f$ for some $t_0 \in \disk$.
We know from the $\lambda$-lemma that $u^t(z) : \disk \times
\Bbb{C} \to \Bbb{C}$ is continuous and thus takes compact sets in
$z$  to compact sets. We also have
$u^t(0) = 0$. Therefore there exists $\epsilon > 0$ such that
$u^t(S_R) \cap \overline{B(0,\epsilon)} = \emptyset$ for all $t$
with $|t| \leq t_0$.

Consider the function $v^t(z) = (1/\epsilon) u^t(\epsilon z)$. For
$|z| = R/\epsilon$ and $|t| \leq |t_0|$, $|v^t(z)| >  (1/\epsilon)
\epsilon = 1$. So the motion of $S_{R/\epsilon}$ under $v^t$ is
disjoint from $S_1$. Also $v^0(z) = z$. The function
$$
g^t(z) = \begin{cases}
z & \text{for } z \in S_1 \\
v^t(z) & \text{for } z \in S_{R/\epsilon}
\end{cases}
$$
is injective and thus a holomorphic motion of $S_1\cup
S_{R/\epsilon}$. The extended $\lambda$-lemma (see Theorem
\ref{th:exlambda}) guarantees an extension of $g^t$ to a
holomorphic motion $G^t : \disk \times \Bbb{C} \to \Bbb{C}$. Note
that $G^t|_{S_1}(z) = z$ and $G^{t_0}(S_{R/\epsilon}) =
S_{R/\epsilon}$. Now $G^{t_0}$ must be modified to map $S_R$ to
itself. Write $z = re^{i \theta}$ and let
$$
w(re^{i \theta}) =
\begin{cases}
re^{i \theta} & \text{for } r \leq 1 \\
(ar + b) e^{i \theta} & \text{for } r \geq 1 \\
\end{cases}
$$
where $a$ and $b$ are defined by $a + b = 1$ and $aR + b =
R/\epsilon$. Note that $w$ is quasiconformal and $w(Re^{i \theta})
= (R/\epsilon) e^{i \theta}$.

We now claim that $F = w^{-1} \circ G^{t_0} \circ w$ is the
desired extension. That is, $F|_{S_1} = \iota$ and $F|_{S_R} = f$.
The map $F$ is quasiconformal since it is the composition of
quasiconformal maps. When $z \in S_1$,
$$
F(z) = w^{-1}(G^{t_0}(w(z))) =  w^{-1}(G^{t_0}(z)) = w^{-1}(z) = z
.
$$
If $z = R e^{i \theta}$ then $u_f(z) = f(z)$ and
\begin{align*}
F(z) & = w^{-1}(G^{t_0}(w(R e^{i \theta}))) \\
     & =  w^{-1}(v^{t_0}((R/\epsilon) e^{i \theta}) \\
     & =  w^{-1}((1/\epsilon) u^{t_0}(\epsilon
           (R/\epsilon) e^{i \theta})) \\
     & =  w^{-1}((1/\epsilon)(u_f(z))) \\
     & =  u_f(z)
\end{align*}
The penultimate equality follows from equalities $|u_f(z)| = R$
and $w^{-1}(Re^{i \theta}/\epsilon ) = R  e^{i \theta}$.
\end{proof}

\begin{remark} The roles of the inner and outer
boundary of the annuli can be interchanged without difficulty.
\end{remark}

\begin{corollary}
\label{co:quasi_annulus_extension} Let $A$ and $B$ be doubly-connected
regions bounded by quasicircles $\alpha_1$ and
$\alpha_2$, and $\beta_1$ and $\beta_2$ respectively. Let $f_1:
\alpha_1 \to \beta_1$ and $f_2: \alpha_2 \to \beta_2$ be
quasisymmetric maps. There exists a quasiconformal map $F : A \to
B$ extending $f_1$ and $f_2$. That is, $F|_{\gamma_1} = f_1$ and
$F|_{\gamma_2} = f_2$.
\end{corollary}
\begin{proof}
Choose numbers $R_1$ and $R_2$ and quasiconformal maps $g_1 : A
\to \ann_{R_1}^{R_2}$ and $g_2 : B \to \ann_{R_1}^{R_2}$ Pick $R'$
such that $R_1 < R' < R_2$.  By applying Lemma
\ref{le:annulus_extension} on $\ann_{R_1}^{R'}$ we get a
quasiconformal map extending $g_2 \circ f_1 \circ g_1^{-1}$ which
is the identity on $S_{R'}$.  Similarly $g_2 \circ f_1 \circ
g_1^{-1}$ can be extended to the identity on $S_{R'}$.  Gluing
these maps and pulling back by $g_1$ and $g_2$ gives the desired
map $F: A \to B$.
\end{proof}

Let $\riem^B$ be a Riemann surface of type $(g,n^-,n^+)$ with boundary
$\partial \riem^B = \partial_1 \riem^B \cup \cdots \cup \partial_{n} \riem^B$ as usual.
\begin{corollary}
\label{co:qs_extension} For $i=1,\ldots,n$, let $f_i : \partial_i
\riem^B \to  \partial_i \riem^B$ be a quasisymmetric self-map of
the $i^{\text{th}}$ boundary component of $\riem^B$.
There exists a
quasiconformal map $f : \riem^B \to \riem^B$ such that
$f|_{\partial_i \riem^B} = f_i$.
\end{corollary}
\begin{proof}
For each $i$, choose $\gamma_i$ to be a quasicircle that is
isotopic to the boundary component $\partial_i \riem^B$.  The
curves $\gamma_i$ can be taken to be mutually disjoint.  Let
$\ann_{\partial_i \riem^B}^{\gamma_i}$ be the annulus bounded by
$\gamma_i$ and $\partial_i \riem^B$. After conformally mapping
$\ann_{\partial_i \riem^B}^{\gamma_i}$ to the plane we can apply
Corollary \ref{co:quasi_annulus_extension} to obtain a
quasiconformal map $F_i : \ann_{\partial_i \riem^B}^{\gamma_i} \to
\ann_{\partial_i \riem^B}^{\gamma_i}$ with $F_i|_{\gamma_i} =
\text{id}$ and $F_i|_{\partial_i \riem^B} = f_i$.  Combining the
$F_i$ and extending by the identity gives the desired extension.
\end{proof}

\begin{corollary} \label{quasicirclemapping}
 Let $\gamma_1$ and $\gamma_2$ be quasicircles in $\disk$ with $0$
 contained in their interiors.  Let
 $f_1:\partial \disk \rightarrow \partial \disk$ and $f_2:
 \gamma_1 \rightarrow \gamma_2$ be quasisymmetric maps. Then there
 exists a quasiconformal map $f:\disk \rightarrow \disk$ such that
 $\left. f\right|_{\partial \disk}=f_1$, $\left. f
 \right|_{\gamma_1}=f_2$, and $f(0) = 0$.
\end{corollary}
\begin{proof}
Choose $\epsilon > 0 $ such that $S_{\epsilon} \cap \gamma_1 =
\emptyset$ and $S_{\epsilon} \cap \gamma_2 = \emptyset$. Apply
Corollary \ref{co:quasi_annulus_extension} to the doubly connected
domains bounded by $S^1$ and $\gamma_1$, and $S^1$ and $\gamma_2$
to get an extension of $f_1$ and $f_2$. Now apply Corollary
\ref{co:quasi_annulus_extension} again to the connected domains
bounded by $\gamma_1$ and $S_{\epsilon}$, and $\gamma_2$ and
$S_{\epsilon}$ to get an extension of $f_2$ and the identity map
on $S_{\epsilon}$. Gluing these two extensions and the identity
map on $B(0,\epsilon)$ produces the desired extension $f$.
\end{proof}
\begin{remark}
 One can extend $f$ to the entire plane.
\end{remark}

\end{subsection}


\begin{subsection}{The correcting map and extension of $\mathrm{PMod}^B(\riem^B)$ to
$\mathrm{PMod}^P(\riem^P)$} \label{se:mcgextensionlemmas}

Given a Riemann surface $\riem^B$ bounded by $n$ closed curves, we
can extend it to a punctured Riemann surface $\riem^P = \riem^B
\#_{\vc{\tau}} (\cdisk_0)^{n}$ by sewing on caps as described in
Section \ref{capsewing}.  In the current section the orientation
of the punctures and boundary components plays no role and can be
safely ignored.

In this section we establish various extension theorems for
quasiconformal mappings.  The main result is the following: any
quasiconformal self-map of $\riem^P$ is isotopic to a
quasiconformal self-map of $\riem^P$ which preserves $\riem^B$. In
fact, the boundary values of the restriction of the new map can be
arbitrarily assigned.  These results are contained in the key
Lemmas \ref{correctingmap} and \ref{strengthenedcorrectingmap},
which will be crucial in establishing the bijection between the
border and puncture models of Teichm\"uller space.

It is also necessary to relate the mapping class group of
$\riem^B$ to that of $\riem^P$.  These results, summarized by
Theorem \ref{longexactsequence}, are present in the literature for
the mapping class groups of diffeomorphisms or homeomorphisms (see
for example \cite[Section 2.1]{HainLoo}, \cite[section 3]{Korkmaz}
and \cite[Section 6]{DicksFormanek}).  For the most part we are
able to apply these known results to our quasiconformal setting.
Where this is not possible we supply short proofs.  These proofs
appear to be new and rely on the extended $\lambda$-lemma in an
essential way.

\begin{lemma} \label{diskhomotopy}
 If $f_1$ and $f_2$ are quasiconformal self-maps of $\cdisk$
 which agree on $\partial \disk$ then
 $f_1$ and $f_2$ are isotopic rel $\partial \disk$.
\end{lemma}
\begin{proof}
 This follows from \cite[Theorem V.1.4]{Lehto}, with the trivial group.
 Alternatively, one can explicitly construct the homotopy as in
 \cite[Theorem IV.3.5]{Lehto}: Let $f_t(z)$ be the point dividing the
 geodesic joining $f_1(z)$ to $f_2(z)$ into segments whose hyperbolic
 lengths have the ratio $t\,:\,1-t$. We note again that homotopies can
 always be replaced with isotopies (see Remark \ref{rm:homo_iso}).
\end{proof}

\begin{remark} \label{diskhomotopywith0}
 If $f_1(0)=0$ and $f_2(0)=0$, one can take the homotopy to
 satisfy $f_t(0)=0$ for all $t$. That is,  $f_1$ and $f_2$ are homotopic
 rel $\partial \disk \cup \{0\}$.  This follows from the explicit
 construction in the proof.
\end{remark}
\begin{corollary} \label{twocurveextension}
 Any $f \in \mathrm{PQC}^B(\riem^B)$ has an extension $\tilde{f} \in
 \mathrm{PQC}^P(\riem^P)$.  Any two such extensions of $f$ are isotopic.
\end{corollary}
\begin{proof}
 Apply Lemma \ref{diskhomotopy} and Remark \ref{diskhomotopywith0} on
 the (punctured) caps with boundary values determined by $f$.
\end{proof}

\begin{proposition}
\label{pr:PB_isotopy} A map $f \in \mathrm{PQC}^B(\riem^B)$ is
isotopic to the identity via an isotopy keeping each boundary
component setwise (but not pointwise) fixed, if and only if any
extension $\tilde{f} \in \mathrm{PQC}^P(\riem^P)$ is isotopic to
the identity on the punctured surface $\riem^P$.
\end{proposition}
\begin{proof}
 This appears in \cite[Proposition 1.3]{Birman}, and references therein,
 for the more general case of homeomorphisms.
\end{proof}

\begin{corollary}
\label{modP_modB_welldefined} The map $\mathrm{PMod}^B(\riem^B)
\to \mathrm{PMod}^P(\riem^P)$ given by taking $[f]$ to an
extension $[\tilde{f}]$ is well defined.
\end{corollary}
Recall that $D_i$ is the $i$th cap on $\riem^P$ and is bounded by
$\partial_i \riem^B$ (considered as a curve in $\riem^P$).
\begin{lemma}
 \label{correctingmap}
 Let $f \in \mathrm{PQC}^P(\riem^P)$ and for $i=1,\ldots,n$, let $N_i$
 be an open neighborhood of $D_i \cup f(D_i)$.  There exists a
 quasiconformal mapping $\alpha:\riem^P \rightarrow \riem^P$ with the
 following properties:
 \begin{enumerate}
  \item $\alpha$ is the identity outside $\cup N_i$,
  \item $\alpha$ takes the curves $\partial_i \riem^B$ to the
   curves $f(\partial_i \riem^B)$, and
  \item $\alpha$ is homotopic to the identity.
 \end{enumerate}
\end{lemma}
\begin{proof}
 We want to construct a separate homotopy on an open neighborhood of
 $\partial_i \riem^B \cup f(\partial_i \riem^B)$ for each $i$, but
 unfortunately the
 curves $f(\partial_i \riem^B)$ may intersect $\partial_j \riem^B$ for
 $i \neq j$.  So the first step is to produce a map that separates
 $\partial_i \riem^B$ from $f(\partial_j \riem^B)$ for $i\neq j$.
 Since $f$ preserves the punctures, $f(D_i)$ contains only the
 puncture $p_i$, and for each fixed $i$ there exists a neighborhood
 $B_i\subset D_i \cap f(D_i)$ of $p_i$ such that $B_i \cap f(D_j) =
 \emptyset$ for all $j \neq i$.  We choose a punctured domain $V_i$
 such that $D_i \subset V_i \subset N_i$ and $V_i$ maps to a quasidisk
 under a local coordinate. Let $s_i : V_i \to V_i$ be a
 quasiconformal map that is the identity on $\partial V_i$ and shrinks
 $D_i$ so that it lies inside $B_i$ (that is, $s_i (D_i) \subset
 B_i)$. The maps $s_i$ can easily be constructed after mapping $V_i$
 to the plane and constructing a suitable map using Corollary
 \ref{quasicirclemapping}.  By choosing the $V_i$ to be mutually
 disjoint we can glue the $s_i$ with the identity map to produce a
 quasiconformal map $s : \riem^P \to \riem^P$ with the property that
 $s(\partial_i \riem^B) \cap f(\partial_j \riem^B) = \emptyset$ for
 all $i,j$ with $i \neq j$.  The fact that $s$ is quasiconformal on
 $\partial V_i$ follows from Theorem \ref{th:qc_remove}.  It follows
 from Lemma \ref{diskhomotopy} that $s$ is homotopic to the identity.

 By applying $s$ we have now reduced the problem to finding a map
 $\beta : \riem^P \to \riem^P$ that takes the curves $s(\partial_i
 \riem^B)$ to the curves $f(\partial_i \riem^B)$.  For each $i$,
 choose a punctured domain $U_i \subset N_i$ that contains $f(D_i)$ (and thus also $s(D_i)$), and that maps to a quasidisk
 under a local coordinate.
 Moreover we can choose the $U_i$ to be mutually disjoint
 (this is why we needed step one).  Choose conformal mappings $g_i :
 U_i \to \disk_0$.  On each copy of the disk we apply Corollary
 \ref{quasicirclemapping} to obtain a map $\beta_i:\disk_0 \rightarrow
 \disk_0$ which is the identity on $\partial \disk$, and takes
 $g_i(s_i(\partial_i \riem^B))$ to $g_i(f(\partial_i\riem^B))$.  The
 maps $\beta_i$ must be homotopic to the identity rel $\partial \disk$
 by Lemma \ref{diskhomotopy}.  Pulling back under the maps $g_i$ and
 gluing to the identity, we obtain $\beta$. Theorem \ref{th:qc_remove}
 ensures that $\beta$ is quasiconformal across the joins.

 Now $\alpha = \beta \circ s$ is the required map. It is homotopic to
 the identity and equal to the identity outside $U_i \cup V_i \subset
 N_i$.
 \end{proof}
 \begin{corollary}
\label{co:correctingmap}
 Every mapping $f \in \mathrm{PQC}^P(\riem^P)$ is homotopic to a
 mapping $\tilde{f} \in \mathrm{PQC}^P(\riem^P)$ which restricts to an
 element of $\mathrm{PQC}^B(\riem^B)$. That is, $\tilde{f}|_{\riem^B}
 \in \mathrm{PQC}^B(\riem^B)$.
\end{corollary}
\begin{proof}
 Apply Lemma \ref{correctingmap} to obtain $\alpha$. The mapping $\tilde{f} =
\alpha^{-1} \circ f$ is homotopic to $f$ and preserves the
boundary curves of $\riem^B$.
\end{proof}

\begin{lemma}[Strengthening of Lemma \ref{correctingmap}]
\label{strengthenedcorrectingmap}
For $i=1,\ldots n$, let $h_i : \partial_i \riem^B \to \partial_i
 \riem^B$ be a quasisymmetric self-map of the boundary component
 $\partial_i \riem^B$.  If $g \in \mathrm{PQC}^P(\riem^P)$ then
there exists $\alpha \in \mathrm{PQC}_0^P(\riem^P)$ such that
the map $g' = \alpha \circ g$ preserves
 $\partial_i \riem^B$ (that is, $g'$ restricts to an element of
 $\mathrm{PQC}^B(\riem^B)$) and $g'|_{\partial_i\riem^B} = h_i$ for each
 $i$.
\end{lemma}
\begin{proof}
 Rather than refining the proof of Lemma \ref{correctingmap} we
instead use Lemma \ref{correctingmap} directly, followed by an
application of Corollary \ref{quasicirclemapping} to obtain the
specified boundary values.

Applying Lemma \ref{correctingmap} with $f = g$, we obtain a map
$\alpha_0 \in \mathrm{PQC}_0^P(\riem^P)$ such that
$\alpha_0(\partial_i \riem^B) = g(\partial_i
\riem^B)$. The quasiconformal map $g_0 = \alpha_0^{-1} \circ g$ is
homotopic to $g$ and preserves the boundaries. Let $U_i \subset
\riem^P$
 be a mutually disjoint
 collection of punctured quasidisks containing $D_i$ (i.e.,  $U_i$ map to quasidisks
 under a local biholomorphic chart.)
 Let $G_i: U_i \rightarrow \disk_0$ be biholomorphic (or quasiconformal) mappings .
 By Corollary \ref{quasicirclemapping}
 there is a mapping $\beta_i: \disk \rightarrow \disk$ such that
 $\left. \beta_i \right|_{S^1}=\text{id}$, $\beta_i(0)=0$, and
 \[
 \beta_i |_{G_i(\partial_i\riem^B)} = G_i \circ g_0
 \circ h_i^{-1} \circ G_i^{-1};  \]
 in particular, $\beta_i$ maps the quasicircle $G_i(\partial_i
 \riem^B)$ to itself.  By Lemma
 \ref{diskhomotopy} and Remark \ref{diskhomotopywith0}, $\beta_i$
 is homotopic to the identity rel $\partial \disk \cup \{0\}$.
 Set
 $$
 \alpha_1 =
 \begin{cases}
      G_i^{-1} \circ \beta_i^{-1} \circ G_i &
\text{on } \overline{U}_i, \quad i=1,\ldots,n \\
      \text{id} & \text{on }(\cup_i U)^c
\end{cases}
$$
and note that $\alpha_1$ is homotopic to the identity and is
quasiconformal on $\partial U_i$ by Theorem \ref{th:qc_remove}.
Let $\alpha = \alpha_1 \circ \alpha_0^{-1}$ and check that
on $\partial_i \riem^B$,
$$
\alpha \circ g = \left( G_i^{-1} \circ ( G_i \circ h_i \circ
g_0^{-1} \circ G_i^{-1} ) \circ G_i \right) \circ g_0 = h_i .
$$
So $g' = \alpha \circ g$ is homotopic to $g$ and agrees with $h_i$
on the boundary curves.
\end{proof}

\begin{corollary}
\label{PQCPtoPQCB}
If $g \in \mathrm{PQC}^P(\riem^P)$ then $g$ is isotopic to a map which
is the identity on $\partial \riem^B$.
\end{corollary}
\begin{proof}
Apply Lemma \ref{strengthenedcorrectingmap} with $h$ the identity
map.
\end{proof}
\begin{corollary}
The map $\chi: \pmcgi[\riem^B] \to \mathrm{PMod}^P(\riem^P)$ sending $[f]$
to any extension $[\tilde{f}]$ is surjective.
\end{corollary}
\begin{proof} The map is well defined by Corollary
\ref{modP_modB_welldefined} and surjective by Corollary
\ref{PQCPtoPQCB}.
\end{proof}

\begin{proposition}
\label{pr:Dehntwistlemma} Let $f \in \mathrm{PQC}(\riem^B)$ and
assume $f|_{\partial \riem^B}$ is the identity. An extension
$\tilde{f} \in \mathrm{PQC}(\riem^P)$ is isotopic to the identity
if and only if the isotopy class of $f$ is an element
of $\mathrm{DB}(\riem^B)$.
\end{proposition}

\begin{proof}
This is a special case of \cite[Theorem 4.1(iii)]{ParisRolfsen}.
\end{proof}

\begin{corollary}
The kernel of $\chi$ is $\mathrm{DB}(\riem^B)$.
\end{corollary}

\begin{theorem} \label{longexactsequence}
The sequence
$$
1 \longrightarrow \mathrm{DB}(\riem^B) \longrightarrow
\pmcgi[\riem^B] \stackrel{\chi}{\longrightarrow} \mathrm{PMod}^P(\riem^P) \longrightarrow 1
$$
is exact.
\end{theorem}
\begin{proof}
This follows directly from the above results.
\end{proof}

\end{subsection}

\end{section}


\begin{section}{The Moduli spaces and their complex structures}
\label{ModuliandTeichmuller}

In this section we define two models of the rigged Riemann and
Teichm\"uller moduli spaces.  The two models can be described as
the `puncture' and `border' models. In either picture, the
relevant Riemann moduli space consists of Riemann surfaces
together with a specification of how to sew them together; this is
called the `rigged' Riemann moduli space.  In the border model,
the boundary data or `rigging' consists of a collection of
mappings of the connected components of the border into $S^1$. In
the puncture model, this boundary data takes the form of local
biholomorphic coordinates around distinguished points.  These two
models are described in Section \ref{riggedRiemann}.

We give a generalization of the standard versions of these models.
This generalization is easiest to state in the border model: we
allow the boundary data of the rigging to be quasisymmetric rather
than analytic.  This allows us to show that the familiar
Teichm\"uller space of a bordered Riemann surface covers the
rigged Riemann moduli space.  The complex structure of
Teichm\"uller space then projects down to rigged Riemann moduli
space. This is accomplished in Section \ref{familiarmodel}.

It is an interesting fact that the boundary data is contained in
the standard Teichm\"uller space of a bordered Riemann surface. In
order to connect with the more familiar approach in conformal
field theory, we construct two `rigged' Teichm\"uller spaces
corresponding to the puncture and border model in Section
\ref{riggedTeichmuller}.  The satisfying relation between these
spaces is completed in Section \ref{Teichmullerrelations}.  The
entire picture is summarized in Section \ref{se:house}.


\begin{subsection}{Puncture and border models of the rigged Riemann
moduli space}
\label{riggedRiemann}
\hfill

\smallskip
\noindent \textbf{Puncture model:} As in Section \ref{preliminaries}, let
$\riem^P$ be a Riemann surface of type $(g,n^-,n^+)$
with oriented and ordered punctures $\mathbf{p} =
(p_1,\ldots,p_{n})$.
The non-negative integers $n^-,n^+$ and $g$
are fixed throughout. We need to describe the rigging.

For ease of
language, we temporarily think of marked points instead of punctures.
 For any point $q \in \riem^P$, let
 $\mathcal{O}(q)$ denote the set of germs of mappings which are
 holomorphic maps from a neighborhood of $q$ into a
 neighborhood of $0 \in \Bbb{C}$, mapping $q$ to $0$.
\begin{definition}
\label{de:qclocalcoord} Let $\mathcal{O}^{\disk}_{qc}(q)$ be the
set of $\phi \in \mathcal{O}(q)$ such that $\disk \subset
\operatorname{Im}(\phi)$, $\phi$ is biholomorphic on $\phi^{-1}(\disk)$, and
$\phi$ extends quasiconformally to a neighborhood of
$\phi^{-1}(\disk)$. For the set of punctures $\mathbf{p}$, let
\begin{multline*}
\mathcal{O}^{\disk}_{qc}(\mathbf{p})=
  \left\{ (\phi^1(p_1), \ldots,
\phi^{n}(p_{n})) \in
\mathcal{O}^{\disk}_{qc}(p_1) \times \cdots \times
  \mathcal{O}^{\disk}_{qc}(p_{n}) \ \st \right. \\
\left. (\phi^i)^{-1}(\cdisk) \cap (\phi^j)^{-1}(\cdisk) = \emptyset, \forall i \neq j \right\}
\end{multline*}
 \end{definition}
For clarification we note that $\phi$ and $\phi'$ in
$\mathcal{O}^{\disk}_{qc}(q)$ are equivalent if and only if they
are equal on some neighborhood of $q$ (and thus on
$\phi^{-1}(\cdisk)$). An element
$$ \vc{\phi}(\mathbf{p}) = \left(\phi^1(p_1), \ldots,
\phi^{n}(p_{n})\right) \in \mathcal{O}^{\disk}_{qc}(\mathbf{p})
$$
will be referred to as the
\textit{local coordinates} or \textit{rigging} of $\riem^P$.
 For brevity we will denote the data of the surface,
punctures, and local coordinates by $(\riem^P, \vc{\phi})$. We
will refer to $(\riem^P, \vc{\phi})$ as a \textit{rigged Riemann
surface}.

\begin{remark}  It should be observed that
$\mathcal{O}^{\disk}_{qc}(q)$ is strictly smaller than the set
$\{ \phi \in \mathcal{O}(q) \st \phi^{-1}$ is biholomorphic on $ \disk\}$.
Considering $\phi^{-1}$, a conformal map of the
disk need not have a quasiconformal extension to a neighborhood.
\end{remark}

 We define an equivalence relation on the set,
$\{(\riem^P,\vc{\phi})\}$, of rigged Riemann surfaces of type
$(g,n^-,n^+)$: we say $(\riem[1]^P, \vc{\phi}_1) \sim_P
(\riem[2]^P, \vc{\phi}_2)$ if and only if there exists a
biholomorphism $\sigma :\riem[1]^P \rightarrow \riem[2]^P$ such
that on $\vc{\phi}^{-1}_1(\cdisk_0)$, we have $\vc{\phi}_1 =
\vc{\phi}_2 \circ \sigma$.  That is, for $i = 1 ,\ldots, n$,
$\phi_1^i = \phi_2^i \circ \sigma$ on each domain
$(\phi_1^i)^{-1}(\cdisk_0)$.  Note that this requires $\sigma$ to
take the $i$th puncture of $\riem[1]^P$ to the $i$th puncture of
$\riem[2]^P$. The equivalence class of $(\riem^P, \vc{\phi})$ will
be denoted by $[\riem^P, \vc{\phi}]$.

 \begin{definition}
\label{PunctureModuli}
 The puncture model of the moduli space of
 rigged Riemann surfaces is
   \[ \widetilde{\mathcal{M}}^P(g,n^-,n^+) =
\{ (\riem^P, \vc{\phi}) \} / \sim_P.
    \]
 \end{definition}
\begin{remark} \label{equivalentequivalence}
This equivalence relation can be stated in the seemingly weaker
form that there exists a biholomorphism $\sigma :\riem[1]^P
\setminus \vc{\phi}^{-1}_1(\disk_0) \to \riem[2]^P \setminus
\vc{\phi}^{-1}_2(\disk_0) $ such that on $\vc{\phi}^{-1}_1(S^1)$,
we have $\vc{\phi}_1 = \vc{\phi}_2 \circ \sigma$. This gives the
same relation, since if $(\phi^i_1)^{-1}$ and $(\phi^i_2)^{-1}$
are analytic on $\disk$ for each $i$, then the map $\sigma$
extends to a biholomorphism $\hat{\sigma} : \riem[1] \to
\riem[2]$. Explicitly
$$
\hat{\sigma} =
\begin{cases}
\sigma & \text{on } \riem[1] \setminus \vc{\phi}_1^{-1}(\disk_0) \\
\vc{\phi}_2^{-1} \circ \vc{\phi}_1 & \text{on }
\vc{\phi}_1^{-1}(\cdisk_0).
\end{cases}
$$
This map is well defined because $\vc{\phi}_1 = \vc{\phi}_2 \circ \sigma$.

\begin{remark}
 In this definition we are consciously imitating the `complex analytic'
 model of the universal Teichm\"uller space due to Bers.  His key insight
 was to extend the complex dilatation of a quasiconformal self-map
 of the disk to the entire plane by setting it to zero outside the disk.
\end{remark}

\end{remark}
\textbf{Border model:} As in Section \ref{preliminaries} let
$\riem^B$
 be a Riemann surface of type $(g,n^-,n^+)$. That is, $\partial \riem
 = (\partial_1\riem^B \cup \cdots \cup \partial_{n} \riem^B)$ where $
 n = n^- + n^+$, there are $n^-$ incoming boundary components, $n^+$
 outgoing boundary components, and sewing in $n$ disks would result in
 a compact Riemann surface of genus $g$.  We fix $n^-, n^+$ and $g$
 throughout.

 A \textit{rigging} of $\riem^B$ is an assignment of a
quasisymmetric map $\psi_i : \partial_i \riem^B \to S^1$ for each
boundary component.  Note that according to Definition
\ref{quasisymmetriccircle} a quasisymmetric map is orientation
preserving.  We denote this ordered set of maps concisely by
$\vc{\psi} = \left( \psi^1, \ldots, \psi^n \right)$. The pair
$(\riem^B, \vc{\psi})$ will be called a \textit{rigged Riemann
surface}.  The notation should prevent any confusion with the
puncture case where the same terminology is used.

 \begin{remark}
See Remark \ref{re:CFTorientation} for the relation to the orientation
of parametrizations in conformal field theory.
\end{remark}

We define an equivalence relation on the set $\{(\riem^B,\psi)\}$ of
type $(g,n^-,n^+)$ rigged Riemann surfaces: $(\riem[1]^B, \vc{\psi}_1) \sim_B
(\riem[2]^B, \vc{\psi}_2)$ if and only if there exists a
biholomorphism $\sigma :\riem[1]^B \to \riem[2]^B$ such
that $\vc{\psi}_1 = \vc{\psi}_2 \circ \sigma$.
The equivalence class of $(\riem^P, \vc{\psi})$
will be denoted $[\riem^B, \vc{\psi}]$.

\begin{definition} The border model of the moduli space of
 rigged Riemann surfaces is
 \[ \widetilde{\mathcal{M}}^B(g,n^-,n^+) = \{ (\riem^B, \vc{\psi} )\} / \sim_B.
 \]
\end{definition}

We will sometimes write $\widetilde{\mathcal{M}}^P(g,n^-,n^+)$ and
$\widetilde{\mathcal{M}}^B(g,n^-,n^+)$ as
$\widetilde{\mathcal{M}}^P(\riem^P)$ and
$\widetilde{\mathcal{M}}^B(\riem^B)$ or even
$\widetilde{\mathcal{M}}^P$ and $\widetilde{\mathcal{M}}^B$, where
the type is assumed to be specified by a base space $\riem^B$ or
$\riem^P$.

We now describe how to convert a punctured surface with local
coordinates into a surface with boundary and quasisymmetric
parametrizations and vice versa. This produces a bijection between
$\widetilde{\mathcal{M}}^P(g,n^-,n^+)$ and
$\widetilde{\mathcal{M}}^B(g,n^-,n^+)$; this is the content of
Theorem \ref{MPMBiso}. Recall that the inversion map $J : \Chat
\to \Chat$ is defined by $J(z) = 1/z$.

Consider a rigged surface $(\riem^P, \vc{\phi})$
and for notational convenience let
$$
\vc{\phi}^{-1}(\disk_0) = \bigcup_{i=1}^n (\phi^i)^{-1}(\disk_0).
$$
We form the surface
$\riem^B = \riem^P \setminus \vc{\phi}^{-1}(\disk_0)$
whose $n$ boundary components are
specified to be incoming (respectively, outgoing)
if the corresponding puncture is negatively (respectively, positively)
oriented.
The map
$$
J \circ \phi^i |_{\partial_i \riem^B} : \partial_i \riem^B \longrightarrow S^1
$$
is a quasisymmetric parametrization by Theorem
\ref{qsisboundaryqc}. The map $J$ is needed to correct the
orientation (see Remark \ref{whyJ} below).
So
$$\left(\riem^P \setminus \vc{\phi}^{-1}(\disk_0),
J \circ \vc{\phi}|_{\vc{\phi}^{-1}(S^1)}\right)
$$
is a rigged Riemann surface.

Considering the converse situation, we begin with the rigged
surface $(\riem^B , \vc{\psi})$.  Choose a set of disjoint annular
neighborhoods $\ann_{\partial_i \riem^B}$ of $\partial_i \riem^B$,
$i=1,\ldots, n$ and let $\ann_{\partial \riem^B} = \bigcup_{i=1}^n
\ann_{\partial_i \riem^B}$. Using Theorem \ref{qsisboundaryqc}, we
can choose a quasiconformal extension of $\psi^i$ to
$\ann_{\partial_i \riem^B}$. Let $\vc{\psi}_{\text{ext}}$ be such
an extension of $\vc{\psi}$ to $\ann_{\partial \riem^B}$.  Note
that each annular neighborhood maps to the interior of $S^1$.  As
in Section \ref{capsewing} we use $\vc{\psi}$ to sew disks onto
$\riem^B$ to produce the punctured surface $\riem^P = \riem^B
\#_{\vc{\psi}} (\cdisk_0)^n$. The orientation of a puncture is
determined by whether the corresponding boundary component is
incoming or outgoing.  Recall that the set of caps is denoted by
$\mathbf{D} = (D_1 \cup \cdots \cup D_n)$.  We claim that
\begin{equation}
\label{psitilde}
\tilde{\vc{\psi}} =
\begin{cases}
J \circ \vc{\psi}_{\text{ext}} & \text{on } \ann_{\partial \riem^B} \\
\text{id} & \text{on } \mathbf{D}
\end{cases}
\end{equation}
is quasiconformal and thus $(\riem^P, \tilde{\vc{\psi}})$ is a
rigged Riemann surface. A direct check using the definition of
sewing shows that $\tilde{\vc{\psi}}$ is well defined.  In terms
of local coordinates on $\riem^P$ given by equation
\eqref{zeta_alpha} in Section \ref{sewingoperation}, the image of
$\partial_i \riem^B \subset \riem^P$ is a Jordan curve on $\Chat$.
This curve is guaranteed to be a quasicircle by Theorem
\ref{th:welding}. Thus $\tilde{\vc{\psi}}$ is quasiconformal by
Theorem \ref{th:qc_remove}.

\begin{remark}
\label{whyJ}
The reason for using $J$ in $J \circ \phi^i$ and $J \circ
\tilde{\vc{\psi}}$ in the constructions above
can be seen most clearly in the
case of the sphere.  Let $\riem^P$ be the Riemann sphere with
a puncture at $0$, and consider the local coordinate $\phi =
\text{id}$. Then $\riem^B = \riem^P \setminus \phi^{-1}(\disk_0)$ is
the upper-hemisphere and $\partial \riem^B$ is $S^1$ but with clockwise
orientation.  So $\phi|_{S^1} : \partial \riem^B \to S^1$ is
orientation reversing.
Equivalently, this can be understood in terms of which side of
$S^1$ the extension of the boundary parametrization maps to.
\end{remark}

\begin{theorem}
\label{MPMBiso}
The map $\mathcal{I}: \widetilde{\mathcal{M}}^P(g,n^-,n^+) \rightarrow
 \widetilde{\mathcal{M}}^B(g,n^-,n^+)$ defined by
 $$
\mathcal{I}([\riem^P,\vc{\phi}])=[\riem^P \backslash
 \vc{\phi}^{-1}(\disk),\left. J \circ \vc{\phi}\right|_{\vc{\phi}^{-1}(S^1)} ]
$$  is a bijection.
\end{theorem}

\begin{proof} \ \newline
 \textit{$\mathcal{I}$ is well defined:} Assuming $(
 \riem[1]^P,\vc{\phi}_1 ) \sim (\riem[2]^P,\vc{\phi}_2)$ in
 $\widetilde{\mathcal{M}}^P(g,n^-,n^+)$, there exists a biholomorphism
 $\sigma: \riem[1]^P \rightarrow \riem[1]^P $ such that $\vc{\phi}_1
 =\vc{\phi}_2 \circ \sigma$ on $\vc{\phi}_1^{-1}(\cdisk_0)$.  After
 restricting $\sigma$ to $\riem[1]^P \setminus \vc{\phi}_1(\disk_0)$
 this is exactly the condition that $({\riem[1]^P} \backslash
 \vc{\phi}_1^{-1}(\disk_0), \left.J \circ
 \vc{\phi}_1\right|_{\vc{\phi}_1^{-1} (S^1)})\sim ({\riem[2]^P}
 \backslash \vc{\phi}_2^{-1}(\disk_0), \left.J\circ \vc{\phi}_2
 \right|_{\vc{\phi}_2^{-1} (S^1)} )$ in
 $\widetilde{\mathcal{M}}^B(g,n^-,n^+)$.
 \smallskip

 \noindent \textit{$\mathcal{I}$ is injective:} Assuming
 $\mathcal{I}([{\riem[1]^P},\vc{\phi}_1])
 =\mathcal{I}([{\riem[2]^P},\vc{\phi}_2]), $ there exists a
 biholomorphism $\sigma: {\riem[1]^P} \backslash
 \vc{\phi}_1^{-1}(\disk) \rightarrow {\riem[1]^P}\backslash
 \vc{\phi}_2^{-1}(\disk)$ such that $J \circ \vc{\phi}_1 = J \circ
 \vc{\phi}_2 \circ \sigma$ on $\vc{\phi}_1^{-1}(S^1)$.  Thus by Remark
 \ref{equivalentequivalence}.
 $[\riem[1]^P,\vc{\phi}_1]=[\riem[2]^P,\vc{\phi}_2]$.
 \smallskip

 \noindent \textit{$\mathcal{I}$ is surjective:} Let
 $[\riem^B,\vc{\psi}] \in \widetilde{\mathcal{M}}^B(g,n^-,n^+)$.  By
 the procedure described above, sewing in unit disks with $\vc{\psi}$
 produces a punctured surface $\riem^P = \riem^B \#_{\vc{\psi}}
 (\cdisk_0)^n$ and an element $[\riem^P, \tilde{\vc{\psi}}]$ in
 $\widetilde{\mathcal{M}}^P(g,n^-,n^+)$.  It follows directly that
 $\mathcal{I}([\riem^P, \tilde{\vc{\psi}}]) = [\riem^B,\vc{\psi}]$.
\end{proof}
\end{subsection}


\begin{subsection}{Teichm\"uller space and the complex structure
on the rigged moduli space}
\label{familiarmodel}


The (usual) Teichm\"uller space of bordered Riemann surfaces actually
contains the data of the boundary parametrizations.  See Section
\ref{Teichmuller} for the definition of Teichm\"uller space, its
complex structure, and for the properties of the mapping class group
action.

As subgroups of $\mathrm{PMod}^B(\riem^B)$, the groups
$\pmcgi[\riem^B]$, $\mathrm{DB}(\riem^B)$
 and $\mathrm{DI}(\riem^B)$ also act on $T^B(\riem^B)$.  Each element is a biholomorphism of $T^B(\riem^B)$.
\begin{lemma}
\label{fixed-point-free} The action of $\pmcgi[\riem^B]$ on $T^B(\riem^B)$
is fixed-point free. In particular,
$\mathrm{DI}(\riem^B)$ and $\mathrm{DB}(\riem^B)$ also act
fixed-point freely on $T^B(\riem^B)$.
\end{lemma}
\begin{proof}
Only the group $\pmcgi[\riem^B]$ needs to be considered as the
other two are subgroups.  Assume that $[\rho] \in \pmcgi[\riem^B]$
fixes $[\riem^B,f,\riem[1]^B]$. Then $[\riem^B,f \circ
\rho,\riem[1]^B] = [\riem^B,f,\riem[1]^B]$ and so there exists a
biholomorphism $\sigma : \riem[1]^B \to \riem[1]^B $ such that $
(f \circ \rho)^{-1} \circ \sigma \circ f$ is isotopic to the
identity rel $\partial \riem^B$. Since $\rho$ is the identity on
$\partial \riem^B$ we see that $\sigma$ must be the identity on
$\partial \riem[1]^B$. Therefore $\sigma$ is the identity on
$\riem[1]^B$ and thus $\rho$ is isotopic to the identity. So
$[\rho]$ is the identity in $\pmcgi[\riem^B]$.
\end{proof}

\begin{lemma}
\label{propdisc} The action of $\pmcgi[\riem^B]$ on $T^B(\riem^B)$
is properly discontinuous.  In particular,
$\mathrm{DI}(\riem^B)$ and $\mathrm{DB}(\riem^B)$ also act
properly discontinuously on $T^B(\riem^B)$.
\end{lemma}
\begin{proof}

Since the action of $\pmcgi[\riem^B]$ on $T^B(\riem^B)$
is fixed-point free,
$\pmcgi[\riem^B]$ acts on $T^B(\riem^B)$ properly discontinuously
(see Definition \ref{de:propdisc}) if and only if
$$
[\rho_n] \cdot [\riem^B, \text{id}, \riem^B] \longrightarrow
[\riem^B, \text{id}, \riem^B]
$$
implies that there exists $N \in \Bbb{N}$ such that $[\rho_n] =
 [\text{id}]$ for all $n\geq N$.

 By sewing on tori to the boundary components we will reduce the
 problem to the compact surface case.  Let $Y$ be a genus-one Riemann
 surface with one boundary component.  Let $\riem^Y$ be the Riemann
 surface (without punctures) obtained by sewing copies of $Y$ to the
 boundary components of $\riem^B$. The parametrizations we use for
 sewing are not important here.  Define $i_* : \pmcgi[\riem^B] \to
 \text{PMod}^P(\riem^Y)$ by $i_*([\rho]) = [\tilde{\rho}]$ where
$$
 \tilde{\rho} =
\begin{cases}
\rho & \text{ on } \riem^B \\
\text{id} & \text{ on the copies of } Y
\end{cases}
$$
From \cite[Theorem 4.1]{ParisRolfsen} we know that $i_*$ is
injective. Note that this is not true if we sew on caps rather
than tori.

It follows directly from the definition of the Teichm\"uller metric
(see Definition \ref{de:Teichmullerdistance}) that
$$
\tau_{B} \left( [\riem^B,\rho,\riem^B],
[\riem^B,\text{id},\riem^B] \right) \geq
\tau_{Y} \left( [\riem^Y,\tilde{\rho},\riem^Y],
[\riem^Y,\text{id},\riem^Y] \right)
$$
where $\tau_B$ and $\tau_Y$ are the Teichm\"uller metrics on
$T^B(\riem^B)$ and $T^P(\riem^Y)$ respectively.  To see this observe
that the the equivalence class in the definition of $\tau_Y$ is larger
than in the definition of $\tau_B$.

Let $[\rho_n]$ be a sequence in $\pmcgi[\riem^B]$ such that
\begin{equation}
\label{seq_conv}
[\riem^B,\rho_n, \riem^B] \longrightarrow [\riem^B, \text{id}, \riem^B] .
\end{equation}
That is,
$$
\tau_{B} ( [\riem^B,\rho_n,\riem^B], [\riem^B,\text{id},\riem^B] )
\rightarrow 0 .
$$
The above inequality implies that
$$
\tau_{Y} ( [\riem^B,\tilde{\rho_n},\riem^B],
[\riem^B,\text{id},\riem^B] ) \rightarrow 0.
$$
By Theorem \ref{ModactionTP} we know the action of
$\text{PMod}^P(\riem^Y)$ on $T^P(\riem^Y)$ is properly
discontinuous. Thus there exists $N \in \Bbb{N}$ such that
$[\tilde{\rho_n}]$ is in the stabilizer for $n\geq N$.
From the definition of proper discontinuity we know that the stabilizer
is finite.

By the injectivity of $i_*$ we see that $[\rho_n]$ is in a finite set
for $n\geq N$. Because $\pmcgi[\riem^B]$ acts fixed-point freely, the
convergence in \eqref{seq_conv} implies that $[\rho_n] = [\text{id}]$
for all $n\geq N$.

Therefore $\pmcgi[\riem^B]$ acts properly
discontinuously on $T^B(\riem^B)$.
\end{proof}

Let $\riem^B$ be of type $(g,n^-,n^+)$ and as in Section \ref{capsewing},
we choose a boundary trivialization $\vc{\tau} =
(\tau_1,\ldots,\tau_{n})$ of $\riem^B$. We define the map
\begin{equation}
\label{TBtoMB}
P_{\riem^B}:  T^B(\riem^B)  \longrightarrow  \widetilde{\mathcal{M}}^B(g,n^-,n^+)
\end{equation}
by $[\riem^B,f,\riem[1]^B] \mapsto [\riem[1]^B,\vc{\tau}\circ
f^{-1}]$. The ordering (and signs) of the boundary components of
$\riem[1]^B$ is defined by pushing forward the ordering (and signs) on
$\riem^B$ by $f$.

\begin{theorem} \label{TBcoversMB}
 The mapping $P_{\riem^B}$ induces a bijection
 \[ P_{\riem^B}^* :  \frac{T^B(\riem^B)}{\pmcgi[\riem^B]} \longrightarrow
 \widetilde{\mathcal{M}}^B(g,n^-,n^+) .
 \]
\end{theorem}
\begin{proof}
\ \newline {\it Well defined:} Assume there is an $[h] \in
 \pmcgi[\riem^B]$ such that $[\riem^B,f_1 \circ h^{-1}
 ,\riem[1]^B]=[\riem^B,f_2,\riem[2]^B]$.  Then there is a
 biholomorphism $\sigma:{\riem}_1 \rightarrow {\riem}_2$ such that
 $f_2^{-1} \circ \sigma \circ f_1 \circ h^{-1}$ is isotopic to the
 identity rel $\partial \riem^B$.  In particular $f_2^{-1} \circ \sigma
 = h \circ f_1^{-1} = f_1^{-1}$ when restricted to $\partial
 {\riem}_1$, so $\vc{\tau} \circ f_2^{-1} \circ \sigma = \vc{\tau}
 \circ f_1^{-1}$; that is, $[{\riem}_1^B, \vc{\tau} \circ
 f_1^{-1}]=[{\riem}_2^B, \vc{\tau} \circ f_2^{-1}]$.

 \smallskip \noindent {\it Injective}: Assume that
 $P_{\riem^B}([\riem^B,f_1,\riem[1]^B]) =
 P_{\riem^B}([\riem^B,f_2,\riem[2]^B])$.  Then there exists a
 biholomorphism $\sigma: \riem[1]^B \rightarrow \riem[2]^B$ such that
 $\vc{\tau} \circ f_2^{-1} \circ \sigma = \vc{\tau} \circ f_1^{-1}$ on
 $\partial {\riem}_1$; thus $f_1^{-1} \circ \sigma^{-1} \circ f_2$ is
 the identity on the boundary.  Let $[h]=[f_1^{-1} \circ \sigma^{-1}
 \circ f_2] \in \pmcgi[\riem^B]$; thus
 \[  [h] [\riem^B,f_1,\riem[1]^B] = [\riem^B,f_1 \circ h,
    \riem[1]^B]=[\riem^B,\sigma^{-1} \circ f_2, \riem[1]^B] =
    [\riem^B,f_2,\riem[2]^B].  \]

 \smallskip \noindent {\it Surjective:}
 Let $[\riem[1],\vc{\psi}] \in \widetilde{\mathcal{M}}^B(g,n^-,n^+)$
 and recall that such data includes an ordering of the boundary
 components.  There exists an $f': \riem^B \rightarrow \riem[1]^B$ which
 is quasiconformal but we must modify it such that the ordering of the
 boundary components on $\riem^B$ and $\riem[1]^B$ correspond under
 $f'$.

 We claim that there exists a quasiconformal map $\gamma:
 \riem^B \to \riem^B$ that permutes the boundary components in any
 specified way. Consider the surface obtained by sewing caps onto the
 boundaries of $\riem^B$. For this punctured surface there exists
 a quasiconformal map which permutes two punctures.  To see this,
 note that there exists a conformal mapping of a neighborhood of
 both punctures onto an open neighborhood of the closed unit disk.
 Clearly there exists a quasiconformal mapping which is the
 identity on $\partial \disk$ and switches the punctures.  Sewing
 this map back to $\riem^B$ we obtain a quasiconformal map which
 interchanges the two punctures in question.

 Such a map can be modified to preserve the boundaries by an
 application of Lemma \ref{correctingmap}. A sequence of such swaps
 can produce a map $\gamma$ giving any desired permutation.  We now
 choose $\gamma$ so that $f = f' \circ \gamma : \riem^B \rightarrow
 \riem[1]^B$ preserves the given orderings.

 By Corollary \ref{co:qs_extension}, there exists a quasiconformal map
 $g:\riem[1]^B \rightarrow \riem[1]^B$ with boundary values $g =
 \vc{\psi}^{-1} \circ \vc{\tau} \circ f^{-1}$.  Thus $P_{\riem^B}([\riem^B, g\circ f ,
 \riem[1]^B]) =  [\riem[1]^B,\vc{\psi}]$.
\end{proof}

From Theorem \ref{TBcoversMB}, Proposition \ref{complex_quotient},
and the fact that the group $\pmcgi[\riem^B]$ acts properly
discontinuously and fixed-point freely as a group of
biholomorphisms (see Lemmas \ref{fixed-point-free}, \ref{propdisc}
and \ref{Modaction_holo}), we immediately have:
\begin{theorem}
\label{th:complex_moduli} The rigged moduli space
$\widetilde{\mathcal{M}}^B(g,n^-,n^+)$ is an infinite-dimensional complex
manifold, with complex structure inherited from $T^B(\riem^B)$.
That is, there exists a unique complex structure on $\widetilde{\mathcal{M}}^B(g,n^-,n^+)$
such that $P_{\riem^B}$ is holomorphic. Moreover, $P_{\riem^B}$ possesses local holomorphic sections.
\end{theorem}

At first sight it appears that the complex structure on
$\widetilde{\mathcal{M}}^B(g,n^-,n^+)$ depends on the choice of
base surface $\riem^B$ and its boundary parametrization
$\vc{\tau}$. It is well known in Teichm\"uller theory that the
complex structure on $T^B(\riem)$ is canonical in the sense that
different choices of base surface give rise to biholomorphically
equivalent spaces. This fact enables us to prove the following
theorem.

\begin{theorem}
\label{unique_complex}
The complex structure on $\widetilde{\mathcal{M}}^B(g,n^-,n^+)$
inherited from $T^B(\riem^B)$ is independent of choice of $\riem^B$
and $\vc{\tau}$.
\end{theorem}
\begin{proof}
Consider two base surface $X$ and $Y$ of type $(g,n^-,n^+)$ with
boundary trivializations $\vc{\tau}_X$ and $\vc{\tau}_Y$
respectively. Let $\widetilde{\mathcal{M}}^B_X$ and
$\widetilde{\mathcal{M}}^B_Y$ be the underlying set
$\widetilde{\mathcal{M}}^B(g,n^-,n^+)$ together with complex
manifold structures inherited from $(X, \vc{\tau}_X)$ and $(Y,
\vc{\tau}_Y)$ respectively.  We need to show that the identity map
$\mathrm{id} : \widetilde{\mathcal{M}}^B_X \to
\widetilde{\mathcal{M}}^B_Y$ is biholomorphic.

Using Corollary \ref{co:qs_extension} as in the proof of
surjectivity in Theorem \ref{TBcoversMB}, we produce a
quasiconformal map $g : Y \to X$ such that $\vc{\tau}_X \circ
g|_{\partial Y} = \vc{\tau}_Y$. This quasiconformal map induces an
\textit{allowable bijection} $g_* : Y \to X $ defined by $g_*([X,
f, \riem[1]^B]) = ([Y, f \circ g , \riem[1]^B])$. The map $g_*$ is
a biholomorphism (see for example \cite[page 122 and 186]{Nag}).
It is straightforward to check that the diagram
$$
\xymatrix{
T^B(X) \ar[r]^{g_*} \ar[d]_{P_X} & T^B(Y) \ar[d]^{P_Y} \\
\widetilde{\mathcal{M}}^B_X  \ar[r]^{\text{id}} &
\widetilde{\mathcal{M}}^B_Y
}
$$
commutes. The maps $P_X$ and $P_Y$ are defined in \eqref{TBtoMB}.
By taking local holomorphic sections of $P_X$ and $P_Y$ we
see that the bottom map, $\text{id}$, is biholomorphic. So
$\widetilde{\mathcal{M}}^B_X$ and $\widetilde{\mathcal{M}}^B_Y$
have identical complex structures.
\end{proof}

\end{subsection}


\begin{subsection}{Rigged Teichm\"uller spaces: puncture and border models}
 \label{riggedTeichmuller}
 We define the \textit{rigged Teichm\"uller spaces} corresponding to
 the puncture and border models of the rigged moduli space.

  In the puncture model, we construct a space of rigged, punctured
 Riemann surfaces as follows.  As in Section \ref{preliminaries}, let
 $\riem^P$ be a punctured Riemann surface of type $(g,n^-,n^+)$ with
 oriented punctures $\mathbf{p} = (p_1,\ldots,p_{n})$.  Let
 \[  \overline{M}^P(\riem^P) = \{( \riem^P,f,\riem[1]^P,\vc{\phi})\} , \]
 where $\riem[1]^P$ is another punctured Riemann surface, $f : \riem^P
 \rightarrow \riem[1]^P$ is a quasiconformal mapping, and $\vc{\phi}
 \in \mathcal{O}^{\disk}_{\text{qc}}(\mathbf{p}_1)$ (see Definition
 \ref{de:qclocalcoord}).  Here $\mathbf{p}_1$ denotes the punctures on
 $\riem[1]^P$ with order induced from $\riem^P$ by $f$, that is,
 $p_1^i = f(p^i)$ for $i=1,\ldots,n$.

 We define an equivalence relation on $\overline{M}^P(\riem^P)$ by
 declaring
 \[  (\riem^P,f_1,\riem[1]^P,\vc{\phi}_1) \sim^P
(\riem^P,f_2,\riem[2]^P,\vc{\phi}_2) \]
 in the case that there exists a biholomorphism $\sigma:\riem[1]^P
 \rightarrow \riem[2]^P$ such that $\vc{\phi}_2 \circ \sigma =
 \vc{\phi}_1$ on $\vc{\phi}_1^{-1}(S^1)$ and $f_2^{-1} \circ \sigma
 \circ f_1$ is isotopic to the identity.  Recall that for a punctured
 surface an isotopy necessarily fixes each puncture throughout.
 \begin{remark}
 \label{re:holodisk}
 As in Remark
 \ref{equivalentequivalence}, the condition $\vc{\phi}_2 \circ \sigma
 = \vc{\phi}_1$ on $\vc{\phi}_1^{-1}(S^1)$ is equivalent to having
 equality on $\vc{\phi}_1^{-1}(\cdisk)$. This follows from the fact
 that holomorphic maps are determined by their boundary values.
 \end{remark}

 \begin{definition}
\label{PunctureTeichmuller}
The \textit{rigged Teichm\"uller space} (for punctured Riemann
 surfaces) is
 \[  \widetilde{T}^P(\riem^P)= \overline{M}(\riem^P) / \sim^P.  \]
 \end{definition}
The mapping class group $\mathrm{PMod}^P(\riem^P)$ acts on $\widetilde{T}^P(\riem^P)$
via
$$
[\rho] \cdot [\riem^P,f,\riem[1]^P,\vc{\phi}]
 =[\riem^P,f \circ \rho,\riem[1]^P,\vc{\phi}],
$$
just as in the usual Teichm\"uller space case (see Section
\ref{Teichmuller}). It can be easily shown, as in Lemma
\ref{fixed-point-free}, that this action is fixed-point free.
Later we will show that this action is also properly discontinuous.

 Define
 \begin{equation}
 \label{Pmod}
 P_{\mathrm{mod}} : \widetilde{T}^P(\riem^P) \longrightarrow \widetilde{\mathcal{M}}^P(g,n^-,n^+)
 \end{equation}
 by $P_{\mathrm{mod}} \left( [\riem^P,f,\riem[1]^P, \vc{\phi}_1]\right) = [\riem[1]^P,\vc{\phi}_1]$.
 We immediately have:
 \begin{proposition}
 \label{Pmodquotient}
 The map $P_{\mathrm{mod}}$ induces a bijection
    \[ P_{\mathrm{mod}}^*:  \frac{\widetilde{T}^P(\riem^P)}{\mathrm{PMod}^P(\riem^P)}
    \longrightarrow \widetilde{\mathcal{M}}^P(g,n^-,n^+) .
        \]
 \end{proposition}
 \begin{proof} The map is surjective
 because given any ordering of the punctures on $\riem[1]^P$, there
 exists a quasiconformal map $f: \riem^P \to \riem[1]^P$ that induces
 the given ordering (as in the proof of surjectivity in
 Proposition \ref{TBcoversMB}.)  The fact that the
 map is well-defined and injectivity
 are a simple consequence of the definitions.
 \end{proof}

The border model of rigged Teichm\"uller space is given by a reduced
Teichm\"uller space with boundary data.  Fix a base Riemann surface
$\riem^B$ of a given type $(g,n^-,n^+)$, which thus fixes an
assignment of $\pm$ to each boundary component as well as an ordering
of the set of components. Consider the
space of quadruples $(\riem^B,f_1,\riem[1]^B,\vc{\psi}_1)$ where
$\riem[1]^B$ is another Riemann surface, $f_1: \riem^B \rightarrow
\riem[1]^B$ is a quasiconformal map and $\vc{\psi}_1$ is a set of
quasisymmetric parametrizations of $\partial \riem[1]^B$ (see `Border
model' in Section \ref{riggedRiemann} for details).  The ordering of
the boundary components on $\riem[1]^B$ is induced by $f$.

We say that
\[  (\riem^B,f_1,\riem[1]^B,\vc{\psi}_1) \sim_\#
    (\riem^B,f_2,\riem[2]^B,\vc{\psi}_2) \] if there exists a
biholomorphism $\sigma:\riem[1]^B \rightarrow \riem[2]^B$ such
that $f_2^{-1} \circ \sigma \circ f_1$ is isotopic to the
identity and $\vc{\psi}_2 \circ \sigma = \vc{\psi}_1$.  It is
important to observe that we do not require that the isotopy is
`rel boundary'.
\begin{definition}
 The rigged reduced Teichm\"uller space of a bordered Riemann
 surface $\riem^B$ is
 \[  \rqt(\riem^B) \cong \{ (\riem^B,f_1,\riem[1]^B,\vc{\psi}_1)
     \}/\sim_\#.  \]
\end{definition}

The border and puncture models are equivalent.  This can be seen by
sewing `caps' onto $\riem^B$ and $\riem[1]^B$ to produce an element in
$\widetilde{T}^P(\riem^P)$. We describe this procedure carefully
before giving the proof of equivalence.  The analogous but simpler
process for rigged moduli spaces was covered in Section
\ref{riggedRiemann}. Certain technical points and notation will not be
repeated.

As in Section \ref{capsewing}, let $\vc{\tau}=
(\tau_1,\ldots,\tau_{n})$ be a trivialization of $\partial
\riem^B$, let $\riem^P = \riem^B \#_{\vc{\tau}} (\cdisk_0)^{n}$,
and let $\mathbf{D}=(D_1 \cup \cdots \cup D_n)$ be the images of the
disks in $\riem^P$. Given $(\riem^B,f,\riem[1]^B,\vc{\psi}_1)$ we
now describe a way to construct an element of
$\widetilde{T}^P(\riem^P)$.  Let $\riem[1]^P = \riem[1]^B
\#_{\vc{\psi}_1} (\cdisk_0)^{n}$ be the punctured surface obtained
by sewing on caps using the parametrization $\vc{\psi}_1$. Note
that $\partial_i \riem[1]^B$ and $\partial \cdisk_0$ are
identified by $\text{id} \circ J \circ \psi^i_1$.  For
$i=1,\ldots,n$, the map
$$
h^i = J \circ \psi_1^i \circ f \circ
(\tau^i)^{-1} \circ J : S^1 \to S^1
$$
is quasisymmetric and thus can be
extended to a quasiconformal map $\tilde{h}^i : \cdisk_0 \to \cdisk_0$.
Using these maps we extend $f$ to a quasiconformal map $\tilde{f} :
\riem^P \to \riem[1]^P$ defined by
\begin{equation}
\label{eq:ftilde}
\tilde{f}(x) =
\begin{cases}
f(x) & \text{for } x\in \riem^B \\
\tilde{h}^i(x) & \text{for } x \in D_i
\end{cases}
\end{equation}
The boundary values $h^i$ were chosen precisely to ensure
$\tilde{f}$ is well defined.  By construction, $\tilde{f}$ is
quasiconformal everywhere except possibly on the seams $\partial_i
\riem^B \subset \riem^P$, but Theorem \ref{th:qc_remove} guarantees that
$\tilde{f}$ is in fact quasiconformal everywhere.

Let $\mathbf{D}_1$ be the union of the caps on $\riem[1]^P$ and let
$\ann_{\partial \riem[1]^B}$ be the union of disjoint annular
neighborhoods of the
boundary components of $\riem[1]^B$.
As in \eqref{psitilde}, we use
$\vc{\psi}_1$ to construct the local coordinates
\begin{equation}
\label{eq:param_to_coord}
\tilde{\vc{\psi}}_1 =
\begin{cases}
J \circ \vc{\psi}_{\text{ext}} & \text{on } \ann_{\partial \riem[1]^B} \\
\text{id} & \text{on } \mathbf{D}_1
\end{cases}
\end{equation}
on $\riem[1]^P$.  That is, we first use Theorem
\ref{qsisboundaryqc} to extend $\psi_1^i$ to a quasiconformal map,
$\psi^i_{\text{ext}}$, on an annular neighborhood
$\ann_{\partial_i \riem^B}$. Then $J \circ \psi^i_{\text{ext}}$ is
extended to the caps by the identity map.

We finally have an element $[\riem^P, \tilde{f}, \riem[1]^P,
\tilde{\vc{\psi}}_1] \in \widetilde{T}^P(\riem^P)$ as desired. The
following theorem shows that this procedure defines a
bijection between the border and puncture models of rigged
Teichm\"uller space.

\begin{theorem}
\label{th:puncture_rigged_equiv}
The map
\begin{align*}
  \mathcal{J}: \rqt (\riem^B) & \longrightarrow \widetilde{T}^P(\riem^P) \\
  {[\riem^B,f,\riem[1]^B,\vc{\psi}_1]} & \longmapsto
   [\riem^P,\tilde{f},\riem[1]^P,\tilde{\vc{\psi}}_1]
 \end{align*}
 is a bijection.  Here, $\riem^P$ and $\riem[1]^P$ are as above and
 $\tilde{f}:\riem^P \rightarrow \riem[1]^P$ is any quasiconformal
 extension of $f$ that takes the punctures to the centres of the caps.
 For $i=1,\ldots,n$, $\tilde{\psi}^i_1$ is an extension of $J
 \circ \psi^i_1$ to a quasiconformal mapping that takes a neighborhood
 of the $i$th cap into a neighborhood of $\disk$ and is conformal
 on $(\tilde{\psi}^i_1)^{-1}(\disk)$.
\end{theorem}
\begin{proof}
By the above discussion we know that maps $\tilde{f}$ and
$\tilde{\vc{\psi}}_1$ with the stated properties exist.

 \medskip \noindent {\it $\mathcal{J}$ is well-defined}: It is immediate
 that the image of the map is independent of the choice of
 $\tilde{\vc{\psi}}_1$.  By Lemma \ref{diskhomotopy} and Remark
 \ref{diskhomotopywith0},
 $[\riem^P,\tilde{f},{\riem}_1^P,\tilde{\vc{\psi}}_1]$ is independent of
 the choice of extension $\tilde{f}$.  It remains to check that if
 $[\riem^B,f_1,{\riem}_1^B,\vc{\psi}_1]=
 [\riem^B,f_2,{\riem}_2^B,\vc{\psi}_2]$ then
 $[\riem^P,\tilde{f}_1,{\riem}_1^P,\tilde{\vc{\psi}}_1]=
 [\riem^P,\tilde{f}_2,{\riem}_2^P, \tilde{\vc{\psi}}_2]$.

  Assume that there is a biholomorphism $\sigma : {\riem}_1^B
  \rightarrow {\riem}_2^B$ such that $f_2^{-1} \circ \sigma \circ f_1$
  is isotopic to the identity and $\vc{\psi}_2 \circ \sigma =
  \vc{\psi}_1$ on $\partial \riem^B$.  The second condition implies
  that $\sigma$ extends to a biholomorphism $
  \tilde{\sigma}:{\riem}_1^P \rightarrow {\riem}_2^P$ by setting it to
  the identity on the caps.  Applying Proposition \ref{pr:PB_isotopy}
  to the extension $\tilde{f}_2^{-1} \circ \tilde{\sigma} \circ
  \tilde{f}_1$ of $f_2^{-1} \circ \sigma \circ f_1$, we see that
  $\tilde{f}_2^{-1} \circ \tilde{\sigma} \circ \tilde{f}_1$ is
  isotopic to the identity.  Thus $[\riem^P,
  \tilde{f}_1,\riem[1]^P,\tilde{\vc{\psi}}_1] =
  [\riem^P,\tilde{f}_2,\riem[2]^P, \tilde{\vc{\psi}}_2]$.

  \medskip \noindent {\it $\mathcal{J}$ is injective}: Assume that
 $\mathcal{J}\left([\riem^B,f_1,\riem[1]^B,\vc{\psi}_1]\right) =
 \mathcal{J}\left([\riem^B,f_2,\riem[2]^B,\vc{\psi}_2]\right)$. Then
 there exists a biholomorphism $\sigma: \riem[1]^P \rightarrow
 \riem[2]^P$ such that $\tilde{\vc{\psi}}_2 \circ \sigma =
 \tilde{\vc{\psi}}_1$ on $\tilde{\vc{\psi}}_1^{-1}(S^1)$ (in fact on
 $\tilde{\vc{\psi}}_1^{-1}(\cdisk)$ by Remark \ref{re:holodisk}) and
 $\tilde{f}_2^{-1} \circ \sigma \circ \tilde{f}_1$ is isotopic to the
 identity on the punctured surface $\riem^P$.  Proposition
 \ref{pr:PB_isotopy} applies to $\tilde{f}_2^{-1} \circ \sigma \circ
 \tilde{f}_1$ and we conclude that $f_2^{-1} \circ \sigma \circ f_1$
 is homotopic to the identity (in general not rel boundary) on
 $\riem^B$.  Therefore $[\riem^B,f_1,\riem[1]^B,\vc{\psi}_1] =
 [\riem^B,f_2,\riem[2]^B,\vc{\psi}_2]$ and $\mathcal{J}$ is injective.

  \medskip \noindent {\it $\mathcal{J}$ is surjective}: Let
  $[\riem^P,g,\riem[1]^P,\vc{\phi}_1] \in \widetilde{T}^P(\riem^P)$
  and $\riem[1]^B = \riem[1]^P \setminus \vc{\phi}_1^{-1}(\disk_0)$.
  The essential step is to show that $g$ can be replaced by a map
  $\hat{g}$ such that its restriction to $\riem^B$ maps onto
  $\riem[1]^B$ without changing the equivalence class in
  $\widetilde{T}^P(\riem^P)$. One subtlety to keep in mind is that
  $\riem[1]^B \# (\cdisk_0)^n$ is conformally equivalent,
  but not equal, to $\riem[1]^P$.

 Let $k: \riem^P \to \riem^P$ be a quasiconformal extension of $g^{-1}
 \circ \vc{\phi}_1^{-1} \circ \vc{\tau}|_{\partial \riem^B}$.
 The existence of $k$ is guaranteed by
 Corollary \ref{co:qs_extension}. Let $\alpha$ be the correcting map
 obtained by applying Lemma \ref{correctingmap} with$f=k$, and let $\hat{g} = g \circ
 \alpha$. From the properties of $\alpha$ it follows that $\hat{g}$ is
 homotopic to $g$ and $ \hat{g}(\partial_i \riem^B) =
 (\vc{\phi}_1^i)^{-1}(S^1)$.

 The rest of the proof only involves keeping
 track of the details of the sewing operations.  On
 $\vc{\phi}_1^{-1}(S^1)$ let $\vc{\psi} = 1 / \vc{\phi}$. For each $i$
 let
$$
\tilde{\psi}_i(x) =
\begin{cases}
\psi^i_1(x) & \text{for }  x \in \phi^i_1|_{\riem[1]^B} \\
x & \text{for }  x \in \cdisk_0
\end{cases}
$$ be the local coordinates on $\riem[1]^B \#_{\vc{\psi}_1}
(\cdisk_0)^{n}$. Let $f = \hat{g}|_{\riem^B}$ and $\tilde{f}$ be defined
as in \eqref{eq:ftilde}.  We claim that with this $\tilde{f}$ and
$\tilde{\vc{\psi}}_1$,
$$
\mathcal{J}[\riem^B, f,\riem[1]^B, \vc{\psi}_1] = [\riem^P,g,\riem[1]^P,\vc{\phi}_1]
$$
The equivalence is determined by the biholomorphism
$\tilde{\sigma} :
\riem[1]^P \to \riem[1]^B \#_{\vc{\psi}_1} (\cdisk_0)^{n}$
defined by
$$
\tilde{\sigma} =
\begin{cases}
\text{id} & \text{on } \riem[1]^B \\
\vc{\phi}_1 & \text{on } \vc{\phi}_1^{-1}(\cdisk_0) .
\end{cases}
$$
It is well defined because the sewing using $\vc{\psi}_1$
identifies $x \in \partial_i \riem[1]^B$ with $1/\psi_i(x) =
\phi_i(x)$. Holomorphicity on $\partial\riem[1]^B$ follows from
Theorem \ref{th:qc_remove}. Moreover, $ \tilde{f}^{-1} \circ
\tilde{\sigma} \circ g$ is homotopic to the identity because $g$
is homotopic to $\hat{g}$ and $ \tilde{f}^{-1} \circ
\tilde{\sigma} \circ \hat{g}$ is the identity except on the caps.
\end{proof}

\end{subsection}


\begin{subsection}{Relation between the Teichm\"uller spaces}
 \label{Teichmullerrelations}

We can create a projection map from $T^B(\riem^B)$ onto
$\widetilde{T}^P(\riem^P)$ from two equivalent points of view.

In the first method, we let $[\riem^B,f,\riem[1]^B] \in
T^B(\riem^B)$. Create the base space $\riem^P = \riem^B
\#_{\vc{\tau}} (\cdisk_0)^{n}$ as usual and let $\mathbf{D} = D_1 \cup
\ldots \cup D_{n}$ be the union of the caps. Consider the local
coordinates
$$
\tilde{\tau}_i =
\begin{cases}
J \circ \tau_i & \text{on an annular neighborhood of } \partial_i
\riem^B \\
\text{id} & \text{on } D_i
\end{cases}
$$ on $\riem^P$ as in equation \eqref{eq:param_to_coord}.  The map
$\tau_i \circ f^{-1}|_{\partial_i \riem[1]^B}$ is a quasisymmetric
boundary parametrization of $\partial_i \riem[1]^B$. Let $\riem[1]^P = \riem[1]^B \#_{\vc{\tau} \circ f^{-1}}
(\cdisk_0)^{n}$, and extend $f$ to $\riem^P$ according to
$$
\tilde{f} =
\begin{cases}
f & \text{on } \riem^B \\
\text{id} & \text{on } \mathbf{D}  .
\end{cases}
$$
The projection is then given by $[\riem^B,f,\riem[1]^B] \mapsto
[\riem^P,\tilde{f},\riem[1]^P,\tilde{\vc{\tau}} \circ \tilde{f}^{-1}]$.

In the second method, let $\tilde{f}'$ be a quasiconformal map on
$\riem^P$ whose dilatation agrees with that of $f$ on $\riem^B$
and is $0$ on $\mathbf{D}$.  Let ${\riem[1]'}^P=\tilde{f}'(\riem^P)$.
The projection is given by $[\riem^B,f,\riem[1]^B] \mapsto
[\riem^P,\tilde{f}',{\riem[1]'}^P,\tilde{\vc{\tau}} \circ
(\tilde{f}')^{-1}]$.

It is not hard to check that both of these maps are well-defined.
The biholomorphic map $\sigma = \tilde{f}' \circ \tilde{f}^{-1} :
\riem[1]^P \rightarrow {\riem[1]'}^P$ establishes that
$[\riem^P,\tilde{f}',{\riem[1]'}^P,\tilde{\vc{\tau}} \circ
(\tilde{f}')^{-1}] =
[\riem^P,\tilde{f},\riem[1]^P,\tilde{\vc{\tau}} \circ
\tilde{f}^{-1}]$.

In the following, we will adopt the first method.
Define
\begin{equation}
\label{PDB}
P_{\mathrm{DB}} : T^B(\riem^B) \longrightarrow \widetilde{T}^P(\riem^P)
\end{equation}
by
$P_{\mathrm{DB}}\left(
[\riem^B,f,\riem[1]^B]\right) = [\riem^P,\tilde{f},
 \riem[1]^P,\tilde{\vc{\tau}}  \circ \tilde{f}^{-1}]$.

\begin{theorem}
\label{th:TBTPiso}
The map
$$
P_{\mathrm{DB}}^* :  \frac{T^B(\riem^B)}{\mathrm{DB}(\riem^B)}
\longrightarrow \widetilde{T}^P(\riem^P)
$$
induced by $P_{\mathrm{DB}}$
is a bijection.
\end{theorem}
\begin{proof} \

\smallskip
 \noindent {\it $P_{\mathrm{DB}}^*$ is well-defined}: The choice of quasiconformal extension
in the definition of $\tilde{\vc{\tau}}$ is immaterial as the
equivalence relation in $\widetilde{T}^P(\riem^P)$ only involves
the local coordinates restricted to the caps.

If $[\riem^B, f_1, \riem[1]^B] = [\riem^B, f_2, \riem[2]^B]$ in
$T^B(\riem^B)$, then the biholomorphism $\sigma : \riem[1]^B \to
\riem[2]^B$ extends by the identity to a map $\tilde{\sigma} :
\riem[1]^P \to \riem[2]^P$. By using $\tilde{\sigma}$, a direct
check shows that
$[\riem^P,\tilde{f}_1,\riem[1]^P,\tilde{\vc{\tau}} \circ f_1^{-1}]
= [\riem^P,\tilde{f}_2,\riem[1]^P,\tilde{\vc{\tau}} \circ
f_2^{-1}]$.

 Let $[h] \in \mathrm{DB}(\riem^B)$, so that
 $[h][\riem^B,f,\riem[1]^B]=[\riem^B,f \circ h,\riem[1]^B]$.
 Define $\tilde{h} : \riem^P \to
 \riem^P$ by
 $$
 \tilde{h} =
 \begin{cases}
 h & \text{on } \riem^B \\
 \text{id} & \text{on }  \mathbf{D}  .
 \end{cases}
 $$
 It follows from Proposition \ref{pr:Dehntwistlemma} that $\tilde{h}$ is isotopic to the
 identity.  Clearly \[ P_{\mathrm{DB}}\left([\riem^B,f \circ h,\riem[1]^B]\right) =
 [\riem^P,\tilde{f} \circ \tilde{h},\riem[1]^P,\tilde{\vc{\tau}} \circ
 \tilde{h}^{-1} \circ \tilde{f}^{-1}].  \]  We need to show that
 $[\riem^P,\tilde{f},\riem[1]^P,\tilde{\vc{\tau}} \circ
 f^{-1}]=[\riem^P,\tilde{f} \circ \tilde{h},
 \riem[1]^P,\tilde{\vc{\tau}} \circ \tilde{h}^{-1} \circ
 \tilde{f}^{-1}]$ in $\widetilde{T}^P(\riem^P)$.  Setting
 $\sigma:\riem[1]^P \rightarrow \riem[1]^P$ to be the identity, we see
 that $\tilde{f}^{-1} \circ \sigma \circ \tilde{f}\circ \tilde{h} =
 \tilde{h}$ is isotopic to the identity, and $\tilde{\vc{\tau}} \circ
 f^{-1} = \tilde{\vc{\tau}} \circ h^{-1} \circ f^{-1}$ on $\partial
 \riem[1]^P$ since $h=\text{id}$ on the boundary.

 \smallskip \noindent {\it $P_{\mathrm{DB}}^*$ is injective}: Assume that
 $P_{\mathrm{DB}}\left([\riem^B,f_1,\riem[1]^B] \right) =
P_{\mathrm{DB}}\left([\riem^B,f_2,\riem[2]^B] \right)$; that is, there
exists a biholomorphism
 $\tilde{\sigma}:\riem[1]^P \rightarrow \riem[2]^P$, such that
 $\tilde{f}_2^{-1} \circ \tilde{\sigma} \circ \tilde{f}_1$ is
 isotopic to the identity, and $\tilde{\vc{\tau}} \circ
 \tilde{f}_2^{-1} \circ \sigma = \tilde{\vc{\tau}} \circ
 \tilde{f}_1^{-1}$ on $\partial \riem[1]^B$.  Therefore $\sigma =
 \tilde{\sigma}|_{\riem^B}$ maps $\riem^B$ to itself and $f_2^{-1}
 \circ \sigma \circ f_1 = \text{id}$ on $\partial \riem^B$. By
 Proposition \ref{pr:Dehntwistlemma}, $\tilde{f}_2^{-1} \circ
 \sigma \circ \tilde{f}_1 |_{\riem^B}$ represents an element
 $[h]$ of $\mathrm{DB}(\riem^B)$; thus
 \[  [h][\riem^B,f_2,\riem[2]^B] = [\riem^B, f_2 \circ h
 ,\riem[2]^B]=[\riem^B, \sigma \circ f_1,
 \riem[2]^B]=[\riem^B,f_1,\riem[1]^B].  \]

 \smallskip \noindent {\it $P_{\mathrm{DB}}^*$ is surjective}: Let
$[\riem^P,g,\riem[1]^P,\vc{\phi}_1] \in \widetilde{T}^P(\riem^P)$.
We will produce $[\riem^B,f',\riem[1]^B] \in T^B(\riem^B)$ that
maps to the given element under $P_{\mathrm{DB}}$. As in the proof of
surjectivity in Theorem \ref{th:puncture_rigged_equiv} we need to
modify $g$ such that it preserves the boundary curves. Moreover,
to obtain $\vc{\phi}_1$ we need to specify the boundary values of
modified map. We apply Theorem \ref{th:puncture_rigged_equiv}.

With $\mathcal{J}$ as in Theorem \ref{th:puncture_rigged_equiv}, it is routine to check that
$$P_{\mathrm{DB}}\left([\riem^B,f,\riem[1]^B]\right) =
\mathcal{J}\left([\riem^B,f,\riem[1]^B,\vc{\tau} \circ f^{-1}]\right).
$$
As $\mathcal{J}$ is onto,
$\mathcal{J}\left([\riem^B,f,\riem[1]^B, \vc{\psi}_1] \right)
= [\riem^P,g,\riem[1]^P,\vc{\phi}_1]$ for some $f$ and $\vc{\psi}_1$.
Let $\tilde{f} : \riem^P \to \riem^P$ be an extension
of $f$ as in the definition of $\mathcal{J}$.

On $\riem^P$, we apply Lemma \ref{strengthenedcorrectingmap} with $g =
\text{id}$ and $h_i = f^{-1} \circ (\psi_1^{i})^{-1} \circ \tau_i$
to produce $g': \riem^P \to \riem^P$ which is homotopic to the
identity and equals $h_i$ on $\partial_i \riem^B$.  The map $f' =
\tilde{f} \circ g'$ is homotopic to $\tilde{f}$ and $f'|_{\partial_i
\riem^B} = (\psi_1^{i})^{-1} \circ \tau_i$.  Thus $\vc{\tau} \circ
(f')^{-1} = \vc{\psi}_1$ and
$$
P_{\mathrm{DB}}\left([\riem^B,f'|_{\riem^B}, \riem[1]^B]\right)
 = \mathcal{J}\left([\riem^B,f,\riem[1]^B, \vc{\psi}_1] \right)
 =  [\riem^P,g,\riem[1]^P,\vc{\phi}_1]
$$
as required. The first equality follows from the fact that the
extension of $f'|_{\riem^B}$, in the definition of
$P_{\mathrm{DB}}$, is homotopic to $f'$ and thus is also homotopic
to $\tilde{f}$.
\end{proof}

As a subgroup of $\pmcgi[\riem^B]$, we know that
$\mathrm{DB}(\riem^B)$ acts properly discontinuously and
fixed-point freely as a group of biholomorphisms on $T^B(\riem^B)$
(Lemmas \ref{fixed-point-free}, \ref{propdisc} and
\ref{Modaction_holo}).  As in Theorem \ref{th:complex_moduli},
Proposition \ref{complex_quotient} immediately implies the
following result.
\begin{corollary}
\label{riggedTPcomplex} The puncture model of rigged Teichm\"uller
space, $\widetilde{T}^P(\riem^P)$, inherits an infinite-dimensional
complex manifold structure from
$T^B(\riem^B)$. That is, there is a unique complex structure on
$\widetilde{T}^P(\riem^P)$ such that $P_{\mathrm{DB}}$ is
holomorphic. Moreover, $P_{\mathrm{DB}}$ possesses local
holomorphic sections.
\end{corollary}

\begin{lemma}
\label{PModonriggedTP}
The group $\mathrm{PMod}^P(\riem^P)$ acts on $\widetilde{T}^P(\riem^P)$
by biholomorphisms.
\end{lemma}
\begin{proof}
Let $[\rho]$ be an element of $\mathrm{PMod}^P(\riem^P)$ and recall that
the action is defined by $[\rho] \cdot [\riem^P,f,\riem[1]^P,
\vc{\phi}_1] = [\riem^P,f \circ \rho, \riem[1]^P, \vc{\phi}_1]$. This
induces a bijection $\rho_* : \widetilde{T}^P(\riem^P) \to
\widetilde{T}^P(\riem^P)$.  The claim is that this map is a
biholomorphism. By Corollary \ref{PQCPtoPQCB}, $\rho$ is isotopic to a
map $\rho'$ that is the identity on $\partial \riem$.  In other words,
$[\rho] = [\rho']$ and $\rho'|_{\riem^B}$ represents an element of
$\pmcgi[\riem^B]$. From Lemma \ref{Modaction_holo} we know that
$(\rho'|_{\riem^B})_* : T^B(\riem^B) \to  T^B(\riem^B)$
is a biholomorphism.
Consider the diagram:
$$
\xymatrix{
T^B(\riem^B) \ar[rr]^{(\rho'|_{\riem^B})_*} \ar[d]_{P_{\mathrm{DB}}} & & T^B(\riem^B) \ar[d]^{P_{\mathrm{DB}}} \\
\widetilde{T}^P(\riem^P) \ar[rr]^{\rho_*} && \widetilde{T}^P(\riem^P) \\
}
$$
The diagram commutes because $\widetilde{f \circ \rho'}$ is
isotopic to $\tilde{f} \circ \rho$. Commutativity and the
existence of local holomorphic sections of $P_{\mathrm{DB}}$
implies that $\rho_*$ is a biholomorphism.
\end{proof}

Define
\begin{equation}
\label{PDBh}
P_{\mathrm{DB}}^{\#} : T^B(\riem^B) \longrightarrow \rqt(\riem^B)
\end{equation}
by
$P_{\mathrm{DB}}^{\#} \left([\riem^B, f, \riem[1]^B] \right) = [\riem^B, f, \riem[1]^B, \vc{\tau} \circ f^{-1}]$.
From Theorems \ref{th:puncture_rigged_equiv}, \ref{th:TBTPiso} and the properties of the action of $\mathrm{DB}(\riem^B)$
we can immediately deduce:
\begin{corollary} \label{trandtb}
The map $P_{\mathrm{DB}}^{\#}$ induces a bijection
$$
\frac{T^B(\riem^B)}{\mathrm{DB}(\riem^B)}
\longrightarrow \rqt(\riem^B)
$$
and $\rqt(\riem^B)$ has a unique complex structure such that $P_{\mathrm{DB}}^{\#}$
is holomorphic. Moreover, $P_{\mathrm{DB}}^{\#}$ possesses local holomorphic sections.
\end{corollary}

Recall that the subspace $\mathrm{DI}(\riem^B)$ of
$\pmcgi[\riem^B]$ generated by  ``internal'' Dehn twists (see
Definition\ref{DI}) acts on $\rqt(\riem^B)$ by \[ [\rho] \cdot
[\riem^B,f,\riem[1]^B, \vc{\psi}_1] =
 [\riem^B,f\circ \rho ,\riem[1]^B, \vc{\psi}_1] \] as usual.

 Define
\begin{equation}
\label{PDI}
P_{\mathrm{DI}} : \rqt(\riem^B) \longrightarrow \widetilde{M}^B(g,n^-,n^+)
\end{equation}
by
$P_{\mathrm{DI}}\left([\riem^B, f, \riem[1]^B, \vc{\psi}_1] \right)= [\riem[1]^B, \vc{\psi}_1]$.
\begin{corollary}
\label{TRandMiso}
The map $P_{\mathrm{DI}}$ induces a bijection
$$
\frac{\rqt(\riem^B)}{\mathrm{DI}(\riem^B)} \longrightarrow
\widetilde{M}^B(g,n^-,n^+)
$$
and $\widetilde{M}^B(g,n^-,n^+) $ has a unique complex structure such that $P_{\mathrm{DI}}$
is holomorphic. Moreover, $P_{\mathrm{DI}}$ possesses local holomorphic sections.
\end{corollary}
\begin{proof}
We could directly  prove the bijection along similar lines as Theorem
\ref{TBcoversMB}. However we will make use of our previous work.
First, observe that
$$
\frac{T^B(\riem^B)}{\pmcgi[\riem^B]} \cong
\frac{T^B(\riem^B)/\mathrm{DB}(\riem^B)}{\mathrm{PModI}(\riem^B)/\mathrm{DB}(\riem^B)}
\cong
\frac{T^B(\riem^B)/\mathrm{DB}(\riem^B)}{\mathrm{DI}(\riem^B)}
$$
by Corollary \ref{MCG_DBDI}. The required isomorphism now follows
from Theorem \ref{TBcoversMB} and Corollary \ref{trandtb}. The
action of $\mathrm{DI}(\riem^B)$ is proper discontinuous and
fixed-point free by Lemmas \ref{fixed-point-free} and
\ref{propdisc}, Theorem \ref{TBcoversMB} and Corollary
\ref{trandtb}. The result now follows from Lemma
\ref{Modaction_holo} and Proposition \ref{complex_quotient}.
\end{proof}

\end{subsection}


\begin{subsection}{Assembly of results: the big picture}
\label{se:house}
In this section we provide a conceptually satisfying commutative diagram and a slightly informal theorem which together summarize many of the results in this paper.
Recall that $\riem^B$ is a bordered surface and $\riem^P$ is the corresponding
punctured surface obtained by sewing on caps as described in ``Border model'', Section \ref{riggedRiemann}.

Consider the commutative diagram:
\begin{equation}
\label{house}
\xymatrix@+20pt{
& T^B(\riem^B) \ar[dl]_{P^{\#}_\mathrm{DB}}
       \ar[dr]^{P_\mathrm{DB}}  & \\
\rqt(\riem^B) \ar[d]_{P_\mathrm{DI}} \ar[rr]^{\cong} & &
\widetilde{T}^P(\riem^P)
\ar[d]^{P_\text{mod}} \\
\widetilde{\mathcal{M}}^B(g,n^-,n^+) \ar[rr]^{\cong} & &
\widetilde{\mathcal{M}}^P(g,n^-,n^+)   .}
\end{equation}
The horizontal isomorphisms are given in Theorems
\ref{MPMBiso} and \ref{th:puncture_rigged_equiv}. The projection
maps are defined in  \eqref{TBtoMB}, \eqref{Pmod}, \eqref{PDB},
\eqref{PDBh}  and \eqref{PDI}. Checking commutativity is routine.

Producing such a diagram was one of the goals of this project.
It gives a full relation between the puncture and and border picture
at both the moduli space and Teichm\"uller space levels.

Before formulating a concluding theorem we
need to show that $P_{\mathrm{mod}}$ is holomorphic.

\begin{lemma}
\label{le:Pmod}
The action of $\mathrm{PMod}^P(\riem^P)$ on $\widetilde{T}^P(\riem^P)$
is properly discontinuous and fixed-point free. The projection $P_{\mathrm{mod}}$ is
holomorphic and possesses local holomorphic sections.
\end{lemma}
\begin{proof}
The bijection $\mathcal{J} : \rqt(\riem^B) \to
\widetilde{T}^P(\riem^P)$ from Theorem
\ref{th:puncture_rigged_equiv} is a biholomorphism by the
commutativity of the top triangle in Diagram \eqref{house} and the
existence of local holomorphic sections of $P_{\mathrm{DB}}$
and $P_{\mathrm{DB}}^{\#}$ (Corollaries
\ref{riggedTPcomplex} and \ref{trandtb}). The actions of $\mathrm{DI}$ on
$\rqt (\riem^B)$ and $\mathrm{PMod}^P(\riem^P)$ on
$\widetilde{T}^P(\riem^P)$ can be seen to be equivalent by
directly using the definitions of the actions and the isomorphism
$\mathcal{J}$. The required properties of the action now follow
from Corollary \ref{TRandMiso} and its proof. In Lemma
\ref{PModonriggedTP} we showed that the action of
$\mathrm{PMod}^P(\riem^P)$ is by biholomorphisms. Thus Proposition
\ref{complex_quotient} guarantees  the stated properties of
$P_{\mathrm{mod}}$.
\end{proof}

\begin{theorem}[Summary of results] \hfill
\label{th:resultssummary}
\begin{enumerate}
\item All the spaces in Diagram \eqref{house} are obtained from
$T^B(\riem^B)$ by quotienting by the action of the mapping class
group and certain subgroups. (Proposition \ref{Pmodquotient},
Theorem \ref{th:TBTPiso} and Corollaries \ref{trandtb} and
\ref{TRandMiso}. )
\item These actions are by biholomorphisms and are properly
discontinuous and fixed-point free. (Lemmas \ref{Modaction_holo},
\ref{ModactionTP}, \ref{fixed-point-free}, \ref{propdisc} and
\ref{PModonriggedTP} and Lemma \ref{le:Pmod}.)
\item With the complex structures inherited from $T^B(\riem^B)$, all
the spaces in Diagram \eqref{house} become complex Banach manifolds.
(Proposition \ref{complex_quotient}, Theorem
\ref{th:complex_moduli}, Corollaries \ref{riggedTPcomplex}, \ref{trandtb} and \ref{TRandMiso} and
Lemma \ref{le:Pmod}.)
\item These complex structures are the unique ones that make all the maps holomorphic. All the maps possess local holomorphic sections.
(Proposition \ref{complex_quotient}.)
\item The horizontal bijections become biholomorphisms. (Commutativity of the diagram and existence of local holomorphic section)
\item The complex structures on the moduli spaces are independent of the
 choice of base surface $\riem^B$ and its boundary trivialization $\vc{\tau}$.
 (Theorem \ref{unique_complex}.)
\end{enumerate}
\end{theorem}

\end{subsection}

\end{section}


\begin{section}{Holomorphicity of Sewing}
\label{se:holo_sewing}

As discussed in Section \ref{introduction}, the sewing operation is
the fundamental geometric operation in conformal field
theory. Holomorphicity of this operation is required in the case of
chiral CFTs, or more formally, in the definition of a weakly conformal
field theory.  In Section \ref{sewing} the sewing operation was
defined for the case of quasisymmetric boundary parametrizations.
We now express this operation as a map between  both the rigged
moduli and Teichm\"uller spaces. These sewing maps are shown to be
holomorphic.

One advantage of working with quasisymmetric maps is the
conceptually satisfying  way in which the sewing maps can be
defined and the holomorphicity proved.

 Recall from Section \ref{Teichmuller} that for a
 bordered Riemann surface, $\riem^B$, the Teichm\"uller space
 $T^B(\riem^B)$ is endowed with the standard complex structure through
 the use of the Bers embedding (see \cite[Chapter 3]{Nag} or
 \cite[V.5.]{Lehto}).  With this structure the fundamental
 projection $\Phi_{\riem^B}: L^{\infty}_{(-1,1)}(\riem^B)_1 \rightarrow
 T^B(\riem^B)$ is holomorphic, and has a local holomorphic section in a
 neighborhood of every point.

 Note that the projection $P_{\riem^B}: T^B(\riem^B) \rightarrow
 \widetilde{\mathcal{M}}(g,n^-.n^+)$ from \eqref{TBtoMB} can
 be defined by first projecting
$$
T^B(\riem^B) \to T^B(\riem^B) / \pmcgi[\riem^B]
$$ and following with the isomorphism $[\riem^B,f,\riem[1]^B] \mapsto
 [\riem[1]^B, \vc{\tau} \circ f^{-1}]$ of Theorem \ref{TBcoversMB}.
 It can be checked directly that  $P_{\riem^B} =P_{\mathrm{DI}} \circ
 P_{\mathrm{DB}^{\#}}$.  From Theorem \ref{th:complex_moduli} we
 know that the projection $P_{\riem^B}$ is
 holomorphic and has local holomorphic sections near every
 point. Actually this projection was used to induce the complex
 structure on $\widetilde{\mathcal{M}}(g,n^-.n^+)$.

We now recall the sewing operation described in Section
\ref{sewingoperation}.  To avoid excessive decorations, we change
notation and let $X$ and $Y$ be bordered Riemann surfaces of type
$(g_X, n^-_X, n^+_X)$ and $(g_Y, n^-_Y, n^+_Y)$ respectively where
$n^+_X >0$ and $n^-_Y>0$.  Let $\vc{\tau}_X = (\tau_X^1, \ldots
\tau_X^{n_X})$ and $\vc{\tau}_Y = (\tau_Y^1, \ldots,\tau_Y^{n_Y})$
be (quasisymmetric) boundary trivializations  of $X$ and $Y$
respectively (see Section \ref{capsewing}). Choose $i$ and $j$
such that $\partial_i X$ is an outgoing boundary component and
$\partial_j Y$ is an incoming boundary component. Let $X \#_{ij}
Y$ be the sewn surface obtained by identifying $\partial_i X$ with
$\partial_j Y$ using $(\tau_Y^j)^{-1} \circ J \circ \tau_X^i$.
Since the choice of $i$ and $j$ is fixed throughout we simply
write $X \# Y$ for $X\#_{ij}Y$. Let $\iota_X: X \to X \# Y$ and
$\iota_Y: Y \to X \# Y$ be the inclusion maps. Let $g_{X\#Y} = g_X
+ g_Y$, $n^-_{X\#Y} = n^-_X + n^-_Y - 1$ and $n^+_{X\#Y} = n^+_X +
n^+_Y -1$. The Riemann surface $X\#Y$ of type
$(g_{X\#Y},n^-_{X\#Y}, n^+_{X\#Y} )$ with boundary trivialization
\begin{equation}
\label{tau_sew}
\vc{\tau}_{X\#Y} =
(\tau_X^1,\ldots,\tau_X^{i-1},\tau_Y^{1},\ldots,\tau_Y^{j-1}, \tau_Y^{j+1},\ldots,\tau_Y^{n_Y}, \tau_X^{i+1},\ldots \tau_X^{n_X})
\end{equation}
will be considered as the base surface for the Teichm\"uller space
$T^B(X \# Y)$. There are other ways of ordering the boundary
components but this issue is not important for our purposes.

\begin{remark}
In conformal field theory the self-sewing operation must be considered.
That is, the sewing of two boundary components of a single surface.
Everything in this section can be altered without difficulty to cover this situation.
\end{remark}

We describe three sewing maps: on the level of Beltrami
differentials, the level of Teichm\"uller space, and the level of
rigged moduli space.
\begin{itemize}
\item $\mathcal{S} : L^{\infty}_{(-1,1)}(X)_1 \times L^{\infty}_{(-1,1)}(Y)_1 \to L^{\infty}_{(-1,1)}(X\#Y)_1$ is defined by
$(\mu , \nu) \mapsto \mu \cup \nu$ where
$$
(\mu \cup \nu)(p)  =
\begin{cases}
\mu(p) & \text{if } p \in \iota_X(X) \\
\nu(p) & \text{if } p \in \iota_Y(Y)
\end{cases}
$$
The values of $\mu \cup \nu$ on the seam of $X \# Y$ are not important
as it is a set of measure zero.

\item $\mathcal{S}_T : T^B(X) \times T^B(Y) \to T^B(X\#Y)$ is defined by
$$
\left( [X,f,X_1] , [Y,g,Y_1] \right) \mapsto [X\#Y, f\cup g, X_1
\# Y_1]
$$
where
$$
(f\cup g)(p)  =
\begin{cases}
f(p) & \text{if } p \in \iota_X(X) \\
g(p) & \text{if } p \in \iota_Y(Y)
\end{cases}
$$ and $X_1$ and $Y_1$ are sewn using the boundary parametrizations
$\tau_X^i \circ f^{-1}$ and $\tau_Y^j \circ g^{-1}$.  By the
definition of the sewing operation, the topologies on $X \# Y$ and
$X_1 \# Y_1$ are such that $f \cup g$ is automatically a
homeomorphism.  Since it is quasiconformal on $\iota_X(X)$ and
$\iota_Y(Y)$, Theorem \ref{th:qc_remove} guarantees it is
quasiconformal on $X \# Y$, by an identical argument to the one
preceding Remark \ref{whyJ}. It is straightforward to check that
$\mathcal{S}_T$ is well defined. For example, if $[X,f,X_1] =
[X,f',X_1']$ via the biholomorphism $\sigma : X_1 \to X_1'$, then
$\sigma_{\#} : X_1 \# Y_1 \to X_1'\#Y_1$ defined by
$$
\sigma_{\#} =
\begin{cases}
\sigma & \text{on } X_1 \\
\text{id} & \text{on } Y_1
\end{cases}
$$
gives the equivalence between $(X\#Y,f\cup g, X_1\#Y_1)$
and $(X\#Y,f'\cup g, X_1'\#Y_1)$.

\item $\mathcal{S}_M: \widetilde{M}^B(g_X, n_X^-,n_X^+) \times
\widetilde{M}^B(g_Y, n_Y^-,n_Y^+) \to \widetilde{M}^B(g_{X\#Y},n^-_{X\#Y}, n^+_{X\#Y}) $ is defined by
$$
\left( [X_1,\vc{\psi}_{X_1}], [Y_1, \vc{\psi}_{X_2} ] \right) \mapsto [X_1
\# Y_1, \vc{\psi}]
$$ where $\vc{\psi}$ is the parametrization of the remaining boundary
components which are ordered in a way analogous to
\eqref{tau_sew}. (To be more precise we should write
$\mathcal{S}_M^{ij}$ where $i$ and $j$ label the boundary components
that are sewed.) It is easy to check that $\mathcal{S}_M$ is well
defined by using maps such as $\sigma_{\#}$.
\end{itemize}

\begin{remark} The spaces and maps
$L^{\infty}_{(-1,1)}(X\# Y)$, $T^B(X \# Y)$, $\mathcal{S}$ and
$\mathcal{S}_Y$ depend on the choice of boundary trivializations
$\vc{\tau}_X$ and $\vc{\tau}_Y$. On the other hand $\mathcal{S}_M$
is independent of $\vc{\tau}_X$ and $\vc{\tau}_Y$.
\end{remark}

\begin{remark}
Being able to sew with quasisymmetric boundary identification is
crucial to defining $\mathcal{S}_T$. In the analytic case this is not
possible, because even if $\vc{\tau}_X$ and $\vc{\tau}_Y$ are chosen
to be analytic, there is no natural way to sew $[X,f,X_1]$ and
$[Y,g,Y_1]$. This is because the maps $\vc{\tau}_X \circ f^{-1}$ and
$\vc{\tau}_Y \circ g^{-1}$ are only quasisymmetric.
\end{remark}

 Consider the following diagram which relates the three sewing
 operations.
\begin{equation}
\label{di:sew}
\xymatrix@+13pt{
L^{\infty}_{(-1,1)}(X)_1 \times L^{\infty}_{(-1,1)}(Y)_1
\ar[rr]^{\mathcal{S}} \ar[d]_{(\Phi_X, \Phi_Y)}
  &&  L^{\infty}_{(-1,1)}(X\# Y) \ar[d]^{\Phi_{X\# Y}} \\
T^B(X) \times T^B(Y) \ar[rr]^{\mathcal{S}_T} \ar[d]_{(P_X,P_Y)}
  &&  T^B(X\# Y) \ar[d]^{P_{X\# Y}} \\
\widetilde{\mathcal{M}}^B(g_X, n^-_X,n^+_X) \times \widetilde{\mathcal{M}}^B(g_Y, n^-_Y,n^+_Y)
\ar[rr]^{\mathcal{S}_M}
  &&  \widetilde{\mathcal{M}}^B(g_{X\#Y},n^-_{X\#Y}, n^+_{X\#Y})
}
\end{equation}

\begin{lemma}
\label{sewdiagram_commutes} Diagram \eqref{di:sew} commutes.
\end{lemma}
\begin{proof}
 The commutativity of the upper rectangle is immediate, since
if $\mu = \mu(f)$ and $\nu = \mu(g)$ then $\mu \cup \nu =
\mu(f\cup g)$.

For the lower rectangle, we take an element
$([X,f,X_1],[Y,g,Y_1])$ of $T^B(X)\times T^B(Y)$ and let
$\vc{\psi}_{X_1} = \vc{\tau}_X \circ f^{-1}$ and $\vc{\psi}_{X_2}
= \vc{\tau}_X \circ f^{-1}$.  Going clockwise, the image of
$([X,f,X_1],[Y,g,Y_1])$ under $P_{X\#Y} \circ \mathcal{S}_T$ is
$[X_1 \# X_2, \vc{\tau}_{X\#Y} \circ (f \cup g)^{-1}]$ where the
sewing is performed using the parametrizations $\tau_X^i \circ
f^{-1}$ and $\tau_Y^j \circ g^{-1}$.  Going anti-clockwise the
image of $([X,f,X_1],[Y,g,Y_1])$ under $\mathcal{S}_M \circ
(P_X,P_Y)$ is $[X_1 \# X_2, \vc{\psi}]$, where $\vc{\psi}$ is
formed from $\vc{\psi}_{X_1}$ and $\vc{\psi}_{X_2}$ with the
appropriate ordering of the remaining boundary components. The
sewing is performed using the parametrizations $\psi^i_{X_1}$ and
$\psi^j_{X_2}$.  A direct check using the definitions of
$\vc{\tau}_{X\#Y}$, $f\cup g$ and $\vc{\psi}$ shows that the
clockwise and anti-clockwise images are identical.
\end{proof}

To show holomorphicity of the sewing maps we need the following
general result. See for example Lehto \cite[page 206]{Lehto} or Nag
\cite[page 87]{Nag}.
\begin{lemma}
\label{Banach_holo}
Let $E$ and $F$ be complex Banach spaces and let $U$ be an open subset
of $E$. Let $F^*$ be the (complex) dual space.
A function $f : U \to
F$ is holomorphic if it is continuous and for every $\alpha \in F^*$
and every $(x,e) \in U \times E$, the function $t \mapsto \alpha \circ
f(x + te)$ is a holomorphic function in some neighborhood of the
origin in $\Bbb{C}$.
\end{lemma}

\begin{lemma}
\label{le:S_holo}
The sewing map $\mathcal{S}$ is holomorphic map.
\end{lemma}
\begin{proof}
 The directional derivatives can be computed directly but it is easier to use
 Lemma \ref{Banach_holo}. Continuity is immediate because if
 $||\mu_1-\mu_2||_\infty < \epsilon$, then $||\mu_1 \cup \nu-\mu_2 \cup
 \nu||_\infty < \epsilon$ and similarly for $\nu$.
For arbitrary $(\mu, \nu)$ and $(\lambda_1, \lambda_2)$ in
$L^{\infty}_{(-1,1)}(X)_1 \times L^{\infty}_{(-1,1)}(Y)_1$,
$$
(\alpha \circ \mathcal{S})((\mu,\nu) + t (\lambda_1, \lambda_2))
 = \alpha(\mu \cup \nu) + t \alpha(\lambda_1 \cup \lambda_2) .
$$
So clearly $t \mapsto (\alpha \circ \mathcal{S})((\mu,\nu) + t
(\lambda_1, \lambda_2))$ is holomorphic in $t$.
\end{proof}

\begin{theorem} \label{th:sewingisholo}
The sewing maps $\mathcal{S}_T$ and $\mathcal{S}_M$ are holomorphic.
\end{theorem}
\begin{proof}
Given any point $(P,Q)$ in $T^B(X) \times T^B(Y)$ let $\sigma_X$
and $\sigma_Y$ be local holomorphic sections of $\Phi_X$ and
$\Phi_Y$ near $p$ and $q$ respectively. Their existence is
guaranteed by Theorem \ref{th:localsectionsexistence}.  It follows
that $\sigma=(\sigma_X,\sigma_Y)$ is a local holomorphic section
of $(\Phi_X, \Phi_Y):L^{\infty}_{(-1,1)}(X)_1 \times
L^{\infty}_{(-1,1)}(Y)_1 \longrightarrow T^B(X) \times T^B(Y)$.
Since Diagram \eqref{di:sew} commutes we have that ${\mathcal{S}_T}
= \Phi_{X \# Y} \circ {\mathcal{S}} \circ \sigma$ and so, by Lemma
\ref{le:S_holo}, $\mathcal{S}_T$ is holomorphic.

Similarly let $ \rho=(\rho_X,\rho_Y):
\widetilde{\mathcal{M}}^B(g_X,n^-_X,n^+_X) \times
\widetilde{\mathcal{M}}^B(g_Y,n^-_Y, n^+_Y) \rightarrow T^B(X)
\times T^B(Y)$ be a local holomorphic section in the neighborhood
of any point in
$\widetilde{\mathcal{M}}^B(g_X,n^-_X,n^+_X) \times
\widetilde{\mathcal{M}}^B(g_Y,n^-_Y, n^+_Y)$. Then $\mathcal{S}_M
= P_{X \# Y} \circ \mathcal{S}_T \circ \rho$ is holomorphic.
\end{proof}

Sewing on caps (as in Section \ref{capsewing}) is a special case of the
sewing operation. This results in a holomorphic map
$T^B(\riem^B)  \rightarrow T^P(\riem^P)$, and
in particular we obtain the following.
\begin{corollary}
\label{co:specialsewingcaps}
The map
\begin{align*}
\mathcal{C}: T^B(\riem^B) & \longrightarrow T^P(\riem^P) \\
[\riem^B,f,\riem[1]^B] & \longmapsto [\riem^B \#_{\vc{\tau}}
(\cdisk_0)^n , f \cup \text{id}, \riem[1]^B \#_{\vc{\tau} \circ f^{-1}}(\cdisk_0)^n]
\end{align*}
is holomorphic.
\end{corollary}
\begin{proof}
The sewing map $\mathcal{S}_T: T^B(\riem^B) \times
(T^B(\cdisk_0))^n \to T^B(\riem^B \#_{\vc{\tau}} (\cdisk_0)^n)$ is
holomorphic. Note that $T^B(\riem^B \#_{\vc{\tau}} (\cdisk_0)^n)$
is the finite-dimensional Teichm\"uller space $T^P(\riem^P)$.
To get the desired map, we fix the second entry of the sewing map
to be the `identity', $([\cdisk_0, \text{id}, \cdisk_0])^n$.
Note
that the map depends on the choice of $\vc{\tau}$.
\end{proof}

Not surprisingly this result enables us to show the compatibility of the complex
structures on $\widetilde{T}^P(\riem^P)$ and the usual Teichm\"uller space $T^P(\riem^P)$.
Consider the following diagram whose left-hand side is the right-hand side of
Diagram \eqref{house}.

\begin{equation} \label{garage}
 \xymatrix@+20pt{
 T^B(\riem^B) \ar[dr]_{P_{\mathrm{DB}}}
       \ar@/^1.7pc/[drr]^{\mathcal{C}}  && \\
& \widetilde{T}^P(\riem^P) \ar[d]_{P_{\text{mod}}} \ar[r]^{\mathcal{F}_T}
& T^P(\riem^P)
\ar[d]^{P_{\text{mod}}} \\
& \widetilde{\mathcal{M}}^P(g,n^-,n^+) \ar[r]^{\mathcal{F}_M} &
\mathcal{M}^P(g,n^-,n^+)   .}
\end{equation}
The horizontal maps just forget the rigging. That is,
\[  \mathcal{F}_T([\riem^P,f,\riem[1]^P, \vc{\phi}_1]) = [\riem^P,f,\riem[1]^P] \ \ \
\mbox{and}  \ \ \ \mathcal{F}_M([\riem[1]^P, \vc{\phi}_1]) =
[\riem[1]^P].  \] Commutativity of the diagram can be checked
directly.

\begin{corollary}
\label{forget}
The map $\mathcal{F}_T :  \widetilde{T}^P(\riem^P) \to T^P(\riem^P)$ is holomorphic.
\end{corollary}
\begin{proof}

We know that $\mathcal{C}$ is holomorphic and that
$P_{\mathrm{DB}}$ possesses local holomorphic sections (Corollary
\ref{riggedTPcomplex}). By commutativity of the diagram,
$\mathcal{F}_{T}$ can locally be expressed as a composition of
these maps, and is thus holomorphic.
\end{proof}

\begin{remark}
The same arguments apply to $\mathcal{F}_M$, but one must be
careful as $\mathcal{M}(g,n^-.n^+)$ is not a complex
manifold. This is because the action of $\text{PMod}^P(\riem^P)$
on $T^P(\riem^P)$ is not fixed-point free.
\end{remark}

\end{section}


\begin{section}{Local structure of rigged Teichm\"uller space}
\label{se:localstructure}

Although we have given a complex manifold structure to the rigged
Teichm\"uller and Riemann Moduli spaces we have not described their local structure.
We will focus on $T^B(\riem^B)$ and $\widetilde{T}^P(\riem^P)$.

From Corollary \ref{forget} we have the holomorphic map
$\mathcal{F}_T :  \widetilde{T}^P(\riem^P) \to T^P(\riem^P)$.
The inverse image of a point is isomorphic to the space of local coordinates.
That is
$$
\mathcal{F}_T^{-1}([\riem^P,f,\riem[1]^P]) =
\left\{ [\riem^P,f,\riem[1]^P, \vc{\phi}]  \st \vc{\phi} \in
\mathcal{O}_{\text{qc}}^{\disk}(\mathbf{p}_1)  \right\}
$$ where $\mathbf{p}_1$ is the list of punctures on $\riem[1]^P$.  The
goal is to show that every point $[\riem^P,f,\riem[1]^P, \vc{\phi}_0]
\in \widetilde{T}^P(\riem^P)$ has a neighborhood of the form $\Omega
\times \mathcal{U}$, where $\mathcal{U}$ is a neighborhood of
$\vc{\phi}_0 \in \mathcal{O}_{\text{qc}}^{\disk}$, and $\Omega$ is a
neighborhood of $[\riem^P,f,\riem[1]^P]$ in the finite-dimensional
Teichm\"uller space $T^P(\riem^P)$.

To reach this goal we must show the space $\mathcal{O}^{\disk}_{\text{qc}}$ is a complex manifold.
How the fibers depend on the Riemann surface must also be understood.
We intend to address these problems in a future publication.

However, a key result will be proved in Section \ref{se:holofamilies}: namely,
if a family of riggings depends holomorphically on
a parameter, then the corresponding family of Teichm\"uller space elements
is holomorphic.
This is the content of  Corollary \ref{t_holo}. The same result for
$\widetilde{T}^P(\riem^P)$ then immediately follows, although
for length considerations we have not presented all the details. This
appears in Corollary \ref{r_holo}.

Once these results are established we sketch, in Section \ref{se:compat_thesis},
the relation of the present results to the standard approach to CFT using analytic riggings.


\begin{subsection}{Holomorphic motion of an annulus}

We first state a crucial lemma.

Let $\gamma_1$ and $\gamma_2$ be Jordan curves such that
$\gamma_1$ is contained in the interior of $\gamma_2$.  Consider a
holomorphic motion $\beta_t$ of $\gamma_1$ (see Definition
\ref{holo_motion}), such that $\beta_t(\gamma_1) \cap \gamma_2 =
\emptyset$ for all $t \in \disk$.

Let $A = \gamma_1 \cup \gamma_2$
and define $h_t: A \to \Bbb{C}$ by
$$
h_t(z) =
\begin{cases}
\beta_t(z) & \text{for } z \in \gamma_1 \\
z &  \text{for } z \in \gamma_2 .
\end{cases}
$$
It follows directly that $h_t$ is a holomorphic motion of $A$.

Let $\ann_{\gamma_1}^{\gamma_2}$ be the annulus bounded by $\gamma_1$
and $\gamma_2$.
Applying the extended $\lambda$-lemma (Theorem \ref{th:exlambda})
we obtain the following.
\begin{lemma}
\label{le:ann_motion}
If $\gamma_1$, $\gamma_2$ and  $\beta_t$ are given as above, then there exists a
holomorphic motion $H_t$ of $\ann_{\gamma_1}^{\gamma_2}$
having the properties guaranteed by Theorem \ref{lambda-lemma}.
In particular,
$H_t|_{\gamma_1} = \beta_t$ and $H_t|_{\gamma_2}$ is the identity.
\end{lemma}

\begin{remark}
Although the proof of this lemma is simple it truly relies on the
power of the extended $\lambda$-lemma. Moreover this lemma in one of
the main technical results needed in the proof of the analyticity of
the sewing operation in \cite{Radnell_thesis}.
\end{remark}

\end{subsection}


\begin{subsection}{Holomorphic family of surfaces} \label{se:holofamilies}

We use Lemma \ref{le:ann_motion} to produce a holomorphic family of
surfaces obtained by cutting out holomorphically varying disks.  The
basic idea is to use the quasiconformal map $H_t$ between annuli to
produce a quasiconformal map between Riemann surfaces.

A family of surfaces is produced in the following way. Assume for
simplicity that $\riem^P$ has a single puncture $p$.  Let $t$ be a
complex parameter and let $\phi_t$ be a family of local coordinates in
$\mathcal{O}_{\text{qc}}^{\disk}(p)$ such that for fixed $z$, $t
\mapsto \phi_t(z)$ is a holomorphic function of $t$. We say that
$\phi_t$ is a holomorphic family of local coordinates.

Our aim is to show that $t \mapsto [\riem^P,\text{id},\riem^P,
\phi_t]$ is a holomorphic map from a neighborhood of $0 \in \Bbb{C}$
to $\widetilde{T}^P(\riem^P)$.  We do this by finding a suitable
holomorphic family of elements in $T^B(\riem^B)$ where $\riem^B =
\riem^P \setminus \phi_0^{-1}(\disk)$.

Let $D_t = \phi_t^{-1}(\disk)$, and $\gamma_t = \phi_t^{-1}(S^1)$.
Consider the bordered Riemann surfaces $\riem[t]^B = \riem^P \setminus
D_t$ with (analytic) boundary parametrizations given by $\phi_t$.
Note that here we allow for a boundary parametrization to be
orientation reversing.

By the definition of $\mathcal{O}_{\text{qc}}^{\disk}(p)$, there
exists $r>1$ such that $\phi_0^{-1}$ is quasiconformal on $B(0,r)$.
Let $U = \phi^{-1}(B(0,r))$ and choose a biholomorphic map $G : U \to
\Bbb{C}$.  For $|t|$ sufficiently small, $D_t$ in contained in $U$.
Let $A_t$ be the annular region on $\riem^P$ bounded by $\partial U$
and $\gamma_t$.

The map $\beta_t = G \circ \phi_t^{-1} \circ \phi_0 \circ G^{-1}
|_{G(\gamma_0)}$ is a holomorphic motion of $G(\gamma_0)$. Applying
Lemma \ref{le:ann_motion} we get a holomorphic motion $H_t$ of
$G(A_0)$ such that $H_t|_{G(\gamma_0)} = \beta_t$ and
$H_t|_{G(\partial U)}$ is the identity.

\begin{proposition}
\label{qc_boundary}
For $|t|$ sufficiently small,
the map $F_t : \riem[0]^B \to \riem[t]^B$ defined by
$$
F_t =
\begin{cases}
\mathrm{id} & \textrm{on}\quad  \riem^P \setminus A_0 \\
G^{-1} \circ H_t \circ G & \textrm{on} \quad A_0
\end{cases}
$$
is quasiconformal and is holomorphic in $t$.
\end{proposition}
\begin{remark} As $\riem[0]^B$ and $\riem[t]^B$ are subsets
of $\riem^P$, it
makes sense to talk about the identity map as well as
holomorphicity in $t$.
\end{remark}
\begin{proof}[Proof of Proposition \ref{qc_boundary}.]
We first show that $F_t$ is well defined.  For $x \in \partial U$,
the fact that $H_t$ is the identity on $ G(\partial  U)$ implies
that $G^{-1} \circ H_t \circ G (x) = x$.  Because $H_t(z)$ is
analytic in $t$ for each fixed $z$, and the other maps are
independent of $t$, we see that $F_t(z)$ is analytic in $t$. The
map $F_t(z)$ is quasiconformal on $\riem[0]^B \setminus \partial
U$ because it is defined by a composition of conformal and
quasiconformal maps. Theorem \ref{th:qc_remove} guarantees that
$F_t$ is quasiconformal on $\partial U$.
\end{proof}

Some standard facts about the complex dilation of a composition of
maps lead to the following.

\begin{lemma}
\label{mu_comp_holo}
Let $\{w_t\}_{t \in \disk}$ be a family of quasiconformal
homeomorphisms of $\Bbb{C}$.  If $t \mapsto \mu(w_t(z))$ is
holomorphic and $f: \Bbb{C} \to \Bbb{C}$ is quasiconformal then the
map $\disk \to L^{\infty}_{(-1,1)}(\Bbb{C})_1$ given by $t \mapsto
\mu(w_t(f(z)))$ is holomorphic.
\end{lemma}

\begin{proposition}
\label{c_holo}
For $|t|$ sufficiently small, the complex dilation $\mu({F_t})$ of
$F_t$ is holomorphic in $t$.  That is, the map $t \mapsto \mu(F_t)$ is
holomorphic.
\end{proposition}
\begin{proof} By Theorem \ref{th:holomotion_dilation}, $\mu(H_t)$
is holomorphic in $t$. Apply Lemma \ref{mu_comp_holo} to $\mu(G^{-1}
\circ H_t \circ G) = \mu(H_t \circ G)$.
\end{proof}

From the holomorphicity of the fundamental projection (see
\ref{th:fundamentalprojectionholomorphic}) we get the
corresponding result for Teichm\"uller space.
\begin{corollary}
\label{t_holo}
 For $|t|$ sufficiently small,
the map $t \mapsto [\riem^B, F_t , \riem[t]^B]$
is holomorphic.
\end{corollary}

Consider the base surface $\riem^B = \riem[0]^B$ whose single boundary
 component is parametrized by $\phi_0$. With some work it can be checked
directly that the holomorphic projection $P_{\mathrm{DB}}:
T^B(\riem^B) \to \widetilde{T}^P(\riem^P)$ defined in \eqref{TBtoMB} sends
$[\riem^B,F_t, \riem[t]^B]$ to $ [\riem^P, \text{id} , \riem^P,
\phi_t]$. To see this, a change of base point must be used along
with the fact that the extension of $F_t$ to the punctured surface
is homotopic to the identity. Corollary \ref{t_holo} now
immediately gives the following.

\begin{corollary}
\label{r_holo}
For $|t|$ sufficiently small,
the map $t \mapsto [\riem^P, \mathrm{id} , \riem^P, \phi_t]$
is holomorphic.
\end{corollary}

We briefly recap the results of this section. Given a holomorphic family of local coordinate
$\phi_t$, define a family of surfaces $\riem[t]^B = \riem^B
\setminus \phi_t^{-1}(\disk)$. Using the extended $\lambda$-lemma,
quasiconformal maps $F_t : \riem^B \to \riem[t]^B$ are obtained.
The family of Teichm\"uller space elements
$ [\riem^B, F_t , \riem[t]^B]$ is a holomorphic curve in $T^B(\riem^B)$.
This holomorphic family projects to a holomorphic family
 $ [\riem^P, \text{id} , \riem^P, \phi_t]$ in $\widetilde{T}^P(\riem^P)$.

\end{subsection}


\begin{subsection}{Relation to analytic rigging}
\label{se:compat_thesis}

In the standard approach to conformal field theory (as defined by
Segal in \cite{SegalPublished}), the boundary components of the
Riemann surfaces are parametrized with analytic maps,
which extend to biholomorphic maps of a collared neighborhood of the
boundary. In the puncture model, the equivalent picture is given by
rigging the punctured surfaces with analytic local coordinates. We denote the
corresponding rigged Teichm\"uller space by
$\widetilde{T}^P_{\mathcal{O}}(\riem^P)$.  The complex structure of this
space and the rigged moduli space are known. It
was worked out in detail in the genus-zero case in \cite{Huang} and in
the higher-genus case in \cite{Radnell_thesis}.  In this section we
outline the compatibility between those complex structures and the one
in the current paper.  The precise statement is the following. \newline
\textbf{Claim:}
\textit{The inclusion map
$$
\mathrm{inc} : \widetilde{T}_{\mathcal{O}}^P(\riem^P)
\longrightarrow \widetilde{T}^P(\riem^P)
$$
is holomorphic. }

It would take significant preparation to properly define the
complex structure on $\widetilde{T}_{\mathcal{O}}^P(\riem^P)$, so this section is
not as detailed as the previous ones.  Moreover, the full details of
the relationship we explore here will be included in a future
publication.

The compatibility hinges on showing that cutting out varying
disks using a holomorphic family of (analytic) local coordinates gives a
holomorphic family in Teichm\"uller space. This is a special case of
Corollary \ref{r_holo}.

We  now briefly describe
(following \cite{Radnell_thesis}),
the complex manifold structure  on the Teichm\"uller space of
analytically rigged Riemann surfaces.

Let $\mathcal{O}$ be the complex vector space of all formal series of
the form $ \sum_{i=1}^{\infty}a_n z^n $, which are absolutely convergent
in some neighborhood of $0 \in \Bbb{C}$.  Let $H_{\disk}$ be the
subspace of $\mathcal{O}$ consisting of series that have a radius of
convergence strictly greater than one.  Let $L_{\disk}$ be the
subspace of $H_{\disk}$ consisting of functions that are one-to-one.
These spaces of germs of holomorphic functions can be described as
(LB)-spaces, that is, as inductive limits of complex Banach spaces.
The theory of holomorphic maps on such spaces closely follows that of
Banach spaces. See for example \cite{KM_97}.

We refer the reader to Section \ref{riggedRiemann} for notation
and related ideas.
In analogy with the local coordinates $\mathcal{O}^{\disk}_{qc}(q)$ from
Definition \ref{de:qclocalcoord}, we define
$$ \mathcal{O}^{\disk}(q) = \{ \phi \in \mathcal{O}(q) \st \disk
\subset \operatorname{Im} (\phi) \text{ , $\phi^{-1}$ biholomorphic on a
neighborhood of } \disk\}.
$$
Let
$\mathcal{O}^{\disk}(\mathbf{p})$ be the space of local coordinates
corresponding to $\mathcal{O}^{\disk}_{qc}(\mathbf{p})$ in Definition
\ref{de:qclocalcoord}.  It can be shown that $\mathcal{O}^{\disk}(q)$
and $\mathcal{O}^{\disk}(\mathbf{p})$ are complex (LB)-manifolds
modelled on $L_{\disk}$ and $(L_{\disk})^n$ respectively.

An analytically rigged surface is a pair $(\riem^P, \vc{\phi})$
where $\riem^P$ is of type $(g,n^-,n^+)$,  and $\vc{\phi} \in
\mathcal{O}^{\disk}(\mathbf{p})$ is the set of local coordinates
at the punctures $\mathbf{p} = (p_1,\ldots,p_n)$. The
corresponding rigged Moduli space and rigged Teichmuller space are
defined exactly as in Definitions \ref{PunctureModuli} and
\ref{PunctureTeichmuller}. We use the notation
$\widetilde{T}^P_{\mathcal{O}}(\riem^P)$ to distinguish from the
earlier case.

In $\widetilde{T}^P_{\mathcal{O}}(\riem^P)$, Schiffer variation of complex structure
can be used to separate the ``Teichm\"uller space part'' from the ``local
coordinate part''.
More precisely, $\widetilde{T}^P_{\mathcal{O}}(\riem^P)$ is a complex
manifold  which, in a neighborhood of $[\riem^P,f,\riem[1]^P,\vc{\phi}]$, has charts of the form
\begin{equation}
\label{NtimesU}
S: \Omega \times \mathcal{U} \longrightarrow
\widetilde{T}_{\mathcal{O}}^P(\riem^P)
\end{equation}
where $\mathcal{U}$ is a neighborhood of $\vc{\phi} \in
\mathcal{O}^{\disk}(\mathbf{p}_1)$ and $\Omega$ is a neighborhood
of $[\riem^P,f,\riem[1]^P]$ in $T^P(\riem^P)$. Recall that
$T^P(\riem^P)$ is a complex manifold of dimension $3g-3+n$.

The initial claim of compatibility of the complex structures on
$\widetilde{T}^P_{\mathcal{O}}(\riem^P)$ and
$\widetilde{T}^P(\riem^P)$ therefore reduces essentially to the
compatibility of $\mathcal{O}^{\disk}(\mathbf{p})$ and
$\widetilde{T}^P(\riem^P)$.  A key step in proving this compatibility
is Corollary \ref{r_holo}.  In the current setting this corollary
shows that a holomorphic curve in $\mathcal{O}^{\disk}(\mathbf{p})$
maps to a holomorphic curve in $\widetilde{T}^P(\riem^P)$.

\end{subsection}

\end{section}


\begin{section}{Concluding remarks}
\label{se:conclusion}

We conclude with some observations.

 The first observation is regarding a possible further application
 of these results to conformal field theory.
 It is well known that the Teichm\"uller space $T^B(\riem^B)$ is
 contained in the universal Teichm\"uller space $T(1)=T(\disk)$.
 Thus a representative of every possible topological type
 $(g,n)$ is contained in $T(1)$.
 It was conjectured by Pekonen \cite{Pekonen} (possibly
 following Nag) that the universal
 Teichm\"uller space is the proper arena for the path integral formulation
 of free bosonic string theory, and might be the basis of a
 non-perturbative formulation.  In other words, the `sum over
 topologies' could be accomplished by using the universal
 Teichm\"uller space (or perhaps some
 suitable subspace) as the space of all paths.  The results of
 the present paper suggest that this may be correct.

 On the other hand, the results of the present paper appear to be an
 application of Segal's definition of conformal field theory
 \cite{SegalPublished} to understanding the Teichm\"uller space of a
 bordered Riemann surface.  First, we have shown that $T^B$ is
 `almost' $\tilde{T}^P$ (that is, up to the action of $\mathrm{DB}$.)
 Second, we have provided two intermediate spaces between
 $\tilde{T}^P$ and $\mathcal{M}^P$, namely $\tilde{\mathcal{M}}^P$ and
 $T^P$ (see Diagram \eqref{garage}).

 This can be interpreted in the following way.  Given a bordered
 Riemann surface $\riem^B$, we want to understand its Teichm\"uller
 space by looking at the compactified surface $\riem^P$.  Again
 ignoring the action of $\mathrm{DB}$, we see that $T^B$ contains the
 following extra information not contained in $\mathcal{M}^P$: the
 riggings, which we add in the horizontal direction of Diagram
 \eqref{garage}, and the markings, which we add in the vertical
 direction of Diagram \eqref{garage}.  In some sense, the riggings can
 be regarded as `external' information in that they specify how
 $\riem^B$ sits inside a compact Riemann surface.  The markings can be
 regarded then as containing `internal' information.

 Finally, we have shown that $T^B/\mathrm{DB}$ is fibred over $T^P$
 with fibres $\mathcal{O}^{\disk}_{qc}$, and that the projection is
 holomorphic (Corollary \ref{forget}).  It is thus of interest to
 describe the complex structure of the fibres
 $\mathcal{O}_{\text{qc}}^{\disk}$ and to show that $T^B$ is a
 holomorphic fiber space which is locally biholomorphic to an open
 subset of $T^P \times \mathcal{O}^{\disk}_{\text{qc}}$.  We hope to
 pursue this point in a future publication.

 \end{section}

\begin{section}{Notation}
\label{notation}

Basic notation:
\begin{itemize}
\item $\riem^B$ - Bordered Riemann surface of finite topological
type.
\item $\riem^P$ - Riemann surface with punctures or marked
points.
\item $S_r$- circle or radius $r$ (in $\Bbb{C}$). \item
$B(0,r)$ - disk radius $r$ (in $\Bbb{C}$)
\item $\ann_{r_1}^{r_2}$
- Standard annulus in $\Bbb{C}$ bounded by $S_{r_1}$ and
$S_{r_2}$.
\item $\ann_{C}$ - An annular neighborhood of a
boundary component of $\riem^B$.
\item $S^1$ - unit circle.
\item $\disk$ - Open unit disk.
\item $\cdisk_0 = \cdisk \setminus \{0\}$
- punctured closed unit disk.
\item $\Chat$ - Riemann sphere.
\item $J : \Chat \to \Chat$ is defined by $J(z) = 1/z$.
\end{itemize}

\noindent General setup:
\begin{itemize}
\item $\riem^B$ is the base surface with $n$ boundary components.
\item $\partial \riem^B = \bigcup_{i=1}^{n} \partial_i \riem^B$
where $\partial_i \riem^B$ are the \textit{ordered} boundary
components. Each component is specified as \textit{incoming} or
\textit{outgoing}.
\item $\vc{\tau} = (\tau_1,\ldots,\tau_n)$ are
fixed quasisymmetric parametrizations of the boundaries. A
\textit{base parametrization} for the \textit{base surface}.
\item $\riem^B
\#_{\vc{\psi}} (\cdisk_0)^n $ - sewing in $n$ disks using boundary
parametrizations $(\psi_1,\ldots,\psi_n)$.
\item $\riem^P = \riem^B \#_{\vc{\tau}} (\cdisk_0)^n$ is the punctured
surface obtained by capping the boundaries.
\item $\riem[1]^B \# \riem[2]^B$ - Result of sewing $\riem[1]^B$ and $\riem[2]^B$ along specified boundary components.
\item $\mathcal{O}_{\text{qc}}^{\disk}(p)$ - Rigging data (puncture model). Space of holomorphic coordinates at $p$ with quasiconformal extensions.
\item $T^B(\riem^B)$ - Teichm\"uller space of a bordered surface.
\item $\widetilde{T}^P(\riem^P)$ - Teichm\"uller space of rigged
surfaces (puncture picture).
\item $\rqt(\riem^B)$ - reduced
Teichm\"uller space of rigged surfaces (border pictue).
\item $\widetilde{\mathcal{M}}^B$ - rigged moduli space (border picture).
\item $\widetilde{\mathcal{M}}^P$ - rigged moduli space (puncture
picture).
\item $\mathrm{PMod}^B(\riem^B)$ - pure mapping class
group.
\item $\pmcgi[\riem^B] = \{ [\rho] \in
\mathrm{PMod}(\riem^B) \st \rho|_{\partial \riem^B} = \text{id} \}$.
\item $\mathrm{DB}(\riem^B)$ -  subgroup of $\pmcgi[\riem^B]$
generated by ``boundary'' Dehn twists.
\item $\mathrm{DI}(\riem^B)$ -  subgroup of $\pmcgi[\riem^B]$ generated
by ``internal'' Dehn twists.

\end{itemize}

\end{section}

\begin{section}{Acknowledgments}
The authors thank Jacob Berstein, Juha Heinonen, Aimo Hinkkanen,
Yi-Zhi Huang, Feng Luo and Curtis McMullen for fruitful
discussions. The first author gratefully acknowledges hospitality and
support from the Max-Planck-Institut f\"ur Mathematik, Bonn.
\end{section}

\bibliographystyle{amsplain}
\bibliography{bib_quasi}

\end{document}